\documentclass[12pt]{article}
\usepackage{amssymb,amsmath,mathrsfs}
\usepackage{graphicx,rotate,subfigure}
\usepackage{cite}
\usepackage[colorlinks=true,
            linkcolor=red,
            urlcolor=blue,
            citecolor=blue]{hyperref}

\usepackage{color}
\long\def\rpl#1!!#2!!{\textcolor{red}{#1} \textcolor{blue}{#2}}

\def \order(#1){{\cal O} \left(#1 \right)}

\evensidemargin=\oddsidemargin
\allowdisplaybreaks
\begin{document}


\begin{center}
{\Large \bf $Z \rightarrow b {\bar b}$ in non-minimal Universal Extra Dimensional Model} \\
\vspace*{1cm}  
{\sf  Tapoja Jha\footnote{tapoja.phy@gmail.com}, Anindya Datta\footnote{adphys@caluniv.ac.in}}
\\
\vspace{10pt} {\small }{\em Department of Physics, University of Calcutta,  
92 Acharya Prafulla Chandra Road, \\ Kolkata 700009, India}

\normalsize
\end{center}

\begin{abstract}
We calculate the effective $Z b\bar b$ coupling at one loop level, in the framework of non-minimal Universal Extra Dimensional (nmUED) model.  Non-minimality in Universal Extra Dimensional (UED) framework is realized by adding kinetic and Yukawa terms with arbitrary coefficients to the action  at  boundary points of the extra space like dimension.  A recent estimation of the Standard Model (SM) contribution to $Zb\bar{b}$ coupling at two loop level, points to a $1.2\sigma$ discrepancy between the experimental data and the SM estimate. We compare our calculation with the 
 difference between the SM prediction and the experimental estimation of the above coupling and constrain the parameter space of nmUED. We 
 also review the limit on compactification radius of UED in view of the new theoretical estimation of SM contribution to $Z b\bar{b}$ coupling. For suitable choice of coefficients of boundary-localized terms, 95\% C.L. lower limit on $R^{-1}$ comes out to be in the ballpark of 800 GeV in the framework of nmUED; while in UED, the lower limit on $R^{-1}$ is 350 GeV which is a marginal improvement over an earlier estimate.
\end{abstract}
\noindent PACS No: {\tt 11.10 Kk, 12.60.-i, 14.70.Hp, 14.80.Rt}

\bigskip
\section{Introduction}
Extra Dimensional theories can offer unique solutions to many long standing puzzles of Standard Model (SM) such as gauge coupling unifications \cite{ued_uni} and  fermion mass hierarchy \cite{hamed}. Most importantly they can provide a Dark Matter candidate of the universe \cite{ued_dm}. In this article, we are interested in a particular incarnation of extra dimensional theory referred as Universal Extra Dimensional Model (UED) where all the SM fields can propagate in $4 +1$ dimensional space-time, the extra dimension (say, $y$) being compactified on a circle ($S^{1}$)  of radius $R$ \cite{acd}. The five dimensional action consists of the same fields of SM and would respect the same $SU(3)_c \times SU(2)_L \times U(1)_Y$ gauge symmetry also. $R^{-1}$ is the typical energy scale at which the four dimensional effective theory 
would start to show up the dynamics of Kaluza-Klein (KK) excitations of SM fields. The masses of KK-modes are $m_n^2 = m^2 + \frac{n^2}{R^2}$; where $n$ is an integer, called the KK-number which corresponds to the discretized momentum in the compactified dimension, $y$. $m$ is any mass parameter that has been attributed to the respective five dimensional field. The $n=0$ mode fields in the effective theory could be identified with SM particles. 
  
To generate the correct structure of chiral fermions in SM, one needs to impose some extra symmetry on the action called orbifolding which is nothing but a discrete $Z_{2}$ symmetry : $y \rightarrow -y$. Fields which have zero modes are chosen to be even under this $Z_{2}$ symmetry. There are KK-excitations of other fields which are odd under this transformation. Consequently they cannot have any zero modes. The space of $y$ is called $S^{1}/Z_{2}$ orbifold with effective domain of $y$ being from $0$ to $\pi R$. These two boundary points will be called fixed points of the orbifold.

The KK mass spectrum in UED are highly degenerate, and radiative corrections to KK masses lift this degeneracy \cite{rad_cor_georgi,rad_cor_cheng}. Radiative corrections to masses include finite bulk corrections originated from the compactification and boundary corrections due to the orbifolding. Boundary corrections to masses have logarithmic dependence on the unknown cut-off scale $\Lambda$. Furthermore they are localized at boundary points. In minimal UED (mUED), boundary terms are considered to be vanishing at the cut-off scale $\Lambda$. This is of course a very special assumption and a more general scenario where this assumption has 
been relaxed is called non-minimal UED (nmUED) \cite{nmUED}. We will be interested in a particular non-minimal scenario in which  kinetic and Yukawa terms involving fields  are added to the five dimensional action, at boundary points. Coefficients of boundary-localized kinetic terms (BLKT)  boundary-localized Yukawa terms (BLYT) and can be chosen as free parameters and experimental data can be used to constrain these. 

Various phenomenological aspects of nmUED have been discussed in \cite{flacke,ddrs}. In particular, studies have been made to constrain non-minimality parameters from the perspective of electroweak observables \cite{flacke}, S, T and U parameters \cite{del}, relic density 
\cite{ddrs1} and from the LHC experiment \cite{lhc}.

Precision electroweak variables like $\rho$(T)-parameter, $R_b$ ($Z$ boson decay width to a pair of $b$ quarks normalized to total hadronic decay width), $A_{FB}^b$ (forward-backward asymmetry of $b$ quarks at $Z$ pole) always have played the role of a
guiding light in search of the new physics. Incidentally, all of these electroweak precision variables  are very much sensitive to the 
radiative corrections and these quantum corrections themselves are amplified by the large top quark mass. In the same spirit, we would like to investigate how one of the precisely known electroweak variable $R_b$ could constrain the nmUED parameter space. 

Estimation of radiative corrections to the $Zb \bar b$ vertex in UED framework has been done previously in Ref.\cite{oliver_zbb, buras}.
However, introduction of  non-minimality through boundary-localized terms (BLT) would shift masses of KK-excitations in a non-trivial manner from their respective UED values. Moreover, some of couplings involving KK-excitations, in nmUED, are also being modified by some factors which are nontrivial functions of BLT parameters. So our calculation would not be a straight forward rescaling of earlier calculations of $Zb\bar{b}$ vertex done in the context of UED \cite{oliver_zbb, buras}. To our knowledge, this is the first effort to estimate the radiative correction to  $Zb\bar{b}$ interaction in non-minimal UED framework.

The plan of the paper is the following. In the next section, we will derive necessary interactions and vertices in the framework of nmUED with a brief introduction of the model. In section 3, we will present some calculational details. Section 4 will be devoted to numerical results including the constraint on the parameter space of nmUED and also a review of $Z b\bar{b}$ constraints in UED. Finally in section 5 we summarize our results and observations.

\section{A brief review of masses and couplings in nmUED}
In this section we will very briefly review the non-minimal Universal Extra Dimensional Model keeping in mind the necessary masses and couplings which will be used in 
our calculations of effective $Zb\bar{b}$ coupling and we will restrict ourselves to boundary-localized kinetic and Yukawa terms only.
 A more detailed account of the model will be found in Ref.  \cite{nmUED,flacke,ddrs}.

We start our discussion with BLKTs for fermions. The resulting action in five dimension is given by
\begin{eqnarray}
\label{actnquark}
\mathcal{S}_{quark} &=& \int d^4 x \int_{0}^{\pi R} dy \Big[\overline{Q} i\Gamma^{M} \mathcal{D}_{M} Q + r_f \{ \delta(y) + \delta(y-\pi R) \} \overline{Q} i\gamma^{\mu} \mathcal{D}_{\mu} P_L Q \nonumber \\ 
 & & + \overline{U} i\Gamma^{M} \mathcal{D}_{M} U + r_f \{ \delta(y) + \delta(y-\pi R) \} \overline{U} i\gamma^{\mu} \mathcal{D}_{\mu} P_R U \nonumber \\ 
 & & + \overline{D} i\Gamma^{M} \mathcal{D}_{M} D + r_f \{ \delta(y) + \delta(y-\pi R) \} \overline{D} i\gamma^{\mu} \mathcal{D}_{\mu} P_R D\Big], 
\end{eqnarray}
where the four component five dimensional fields can be expressed in terms of two component chiral spinors and their Kaluza-Klein excitations as:
\begin{equation}
\label{fermionexpnsn}
Q_{t,b}(x,y) =   \sum^{\infty}_{n=0} \left( \begin{array}{c} Q_{t,b L}^{n}(x) f_L^n(y) \\ Q_{t,b R}^{n}(x) g_L^n(y)\end{array} \right),
U(x,y) = \sum^{\infty}_{n=0}  \left( \begin{array}{c} U_{L}^{n}(x) f_R^n(y) \\ U_{R}^{n}(x) g_R^n(y) \end{array} \right),
D(x,y) = \sum^{\infty}_{n=0}  \left( \begin{array}{c} D_{L}^{n}(x) f_R^n(y) \\ D_{R}^{n}(x) g_R^n(y) \end{array} \right).
\end{equation} 

 In the above expression (and also in the following) $M,N=0,1,2,3,4$ are five dimensional Lorentz indices, with the metric convention $g_{MN} \equiv {\rm diag}(+,-,-,-,-)$. The covariant derivative is defined as $D_M\equiv\partial_M-i\widetilde{g}W_M^a T^a-i \widetilde{g}^\prime B_M Y$, where $\widetilde{g}$ and $\widetilde{g}^\prime$ are the corresponding five dimensional gauge coupling constants of $SU(2)_L$ and $U(1)_Y$ respectively. $T^a$ and $Y$ are the corresponding generators. $\Gamma_M$ are representations of $4 + 1$ dimensional Clifford algebra with $\Gamma_\mu = \gamma_\mu$; $\Gamma_4 = i \gamma_5$. Since we are dealing with only third generation quark, the compact form of doublet is given as $Q = \left(Q_{t}, Q_{b}\right)^{T}$. Upon compactification and orbifolding, this would give rise to the left-handed doublet consisting of $t_{L}^{0}$ and $b_{L}^{0}$. $U$ and $D$ are  four component fields in five dimension from which $t_{R}^{0}$ and $b_{R}^{0}$ would emerge in the four dimensional effective theory. The $y$ dependent wave functions with appropriate boundary conditions are given by
\begin{eqnarray}
f_{L} = g_{R} = N_{Qn} \left\{ \begin{array}{rl}
                \displaystyle \frac{\cos(M_{Q_{n}} \left (y - \frac{\pi R}{2}\right))}{C_{Q_{n}}}  &\mbox{for $n$ even,}\\
                \displaystyle \frac{{-}\sin(M_{Q_{n}} \left (y - \frac{\pi R}{2}\right))}{S_{Q_{n}}} &\mbox{for $n$ odd,}
                \end{array} \right.
                \label{flgr}
\end{eqnarray}
and
\begin{eqnarray}
g_{L} = f_{R} = N_{Qn} \left\{ \begin{array}{rl}
                \displaystyle \frac{\sin(M_{Q_{n}} \left (y - \frac{\pi R}{2}\right))}{C_{Q_{n}}}  &\mbox{for $n$ even,}\\
                \displaystyle \frac{\cos(M_{Q_{n}} \left (y - \frac{\pi R}{2}\right))}{S_{Q_{n}}} &\mbox{for $n$ odd,}
                \end{array} \right.
\end{eqnarray}
with
\begin{equation}
C_{Q_{n}} = \cos\left( \frac{M_{Q_{n}} \pi R}{2} \right)\,\, , 
\,\,\,\,
S_{Q_{n}} = \sin\left( \frac{M_{Q_{n}} \pi R}{2} \right).
\end{equation}

These wave functions satisfy the orthonormality conditions 
\begin{equation}
\int dy \left[1 + r_{f}\{ \delta(y) + \delta(y - \pi R)\}
\right] ~k^n(y) ~k^m(y) = \delta^{n m} = \int dy ~l^n(y) ~l^m(y),
\end{equation}
where $k^{n}(y)$ can be $f_{L}$ or $g_{R}$ and $l^{n}(y)$ corresponds to $g_{L}$ or $f_{R}$. From the above condition,
\begin{equation}
N_{Qn} = \sqrt{\frac{2}{\pi R}}\left[ \frac{1}{\sqrt{1 + \frac{r_f^2 M_{Qn}^{2}}{4} 
+ \frac{r_f}{\pi R}}}\right].
\label{normalisation}
\end{equation}

The  mass $M_{Qn}$ of the $n$th KK-mode is no longer equal to $n/R$ as in UED, rather it satisfies the following transcendental equations 
\begin{eqnarray}
  r_{f} M_{Qn}= \left\{ \begin{array}{rl}
         -2\tan \left(\frac{M_{Qn}\pi R}{2}\right) &\mbox{for $n$ even,}\\
          2\cot \left(\frac{M_{Qn}\pi R}{2}\right) &\mbox{for $n$ odd.}
          \end{array} \right.   
          \label{fermion_mass}      
 \end{eqnarray}
It is evident that, for zero modes $(n=0)$, $M_{Qn}$ vanishes identically. 

The other required couplings of the theory must be supplemented by the action of gauge fields, Higgs field and the Yukawa interaction between  the Higgs and fermions:

\begin{eqnarray}
\label{gauge}
\mathcal{S}_{A} &=& \int d^4 x \int_{0}^{\pi R} dy\Big[-\frac{1}{4}F^{MNa}F_{MN}^{a} -\frac{r_{g}}{4}\{ \delta(y) + \delta(y - \pi R)\}F^{\mu \nu a}F_{\mu \nu}^{a} \nonumber\\
 & & -\frac{1}{4}B^{MN}B_{MN} -\frac{r_{g}}{4}\{ \delta(y) + \delta(y - \pi R)\}B^{\mu \nu }B_{\mu \nu} \Big],\\
\label{higgs}
\mathcal{S}_{\Phi} &=& \int d^4 x \int_{0}^{\pi R} dy \Big[\left(D^{M}\Phi\right)^{\dagger}\left(D_{M}\Phi\right) + r_{\phi}\{ \delta(y) + \delta(y - \pi R)\}\left(D^{\mu}\Phi\right)^{\dagger}\left(D_{\mu}\Phi\right) \Big], \\
\label{yukawa}
\mathcal{S}_{Y} &=& -\int d^4 x \int_{0}^{\pi R} dy \Big[\tilde{y}_{t}\;\bar{Q}\Phi^{c}U + \tilde{y}_{b}\bar{Q}\Phi D \nonumber \\
 & &  +r_y \;\{ \delta(y) + \delta(y-\pi R) \}\{\tilde{y}_{t}\bar{Q}_{L}\Phi^{c}U_{R} + \tilde{y}_{b}\bar{Q}_{L}\Phi D_{R}\}  +\textrm{h.c.}\Big].
\end{eqnarray}

 In the above, $F_{MN}^a \equiv (\partial_M W_N^a - \partial_N W_M^a+\widetilde{g}f^{abc}W_M^bW_N^c)$ is the field strength associated with the $SU(2)_L$ gauge group ($a$ is the $SU(2)$ gauge index) and , $B_{MN}=\partial_M B_N - \partial_N B_M$ is that of the $U(1)_Y$ group.  $\Phi$ and $\Phi^c (\equiv i\tau^2 \Phi^\ast)$ are the standard Higgs doublet and its charge conjugated field; $r_{\phi}$, $r_{g}$ are  are  BLKT parameters for the scalar and gauge fields while $r_y$ is the coefficient for boundary-localized Yukawa interactions respectively. Five dimensional gauge couplings $\tilde g$ and $\tilde g'$  are connected to their four dimensional counterparts via the following relation:
 \begin{equation}
g \; (g') = \frac {\widetilde {g} \;(\widetilde {g'})}{\sqrt{r_{g} + \pi R}}.
\label{gauge_coup}
\end{equation}

In the limit, $r_g = r_{\phi}$\footnote{In general when Higgs and gauge BLKTs are unequal, the differential equation governing the dynamics of gauge profile in $y$ direction
contains a term proportional to $r_{\phi}$ due to breakdown of electroweak symmetry \cite{avirup-gauge-fix} and solutions will be different from those given in eq.\ref{flgr}. } (which we will be using throughout our analysis), gauge and scalar fields have the same $y$ dependent profile as $f_L$ (and $g_R$) given in eq.\ref{flgr}
 and their KK-excitations have masses   $M_{gn}$ ($ = M_{\Phi n}$)  which follow from the same transcendental equation given in eq.\ref{fermion_mass} with $r_f$ 
replaced by $r_g$ ($= r_{\phi}$). $\widetilde{y}_t$ and $\widetilde{y}_b$ denote the Yukawa interactions strengths  for the third generation quarks in the five dimensional theory. 

Finally, one must note down the gauge fixing action, which is very crucial for the calculation at our dispense, as we would proceed 
with our calculation in
't-Hooft Feynman gauge. Following Ref.\cite{avirup-gauge-fix} one can have,

\begin{eqnarray}
\label{gauge_fix}
\mathcal{S}_{\rm GF}^{W} &=& -\frac{1}{\xi _y}\int d^{4}x\int_{0}^{\pi R} \Big\vert\partial_{\mu}W^{\mu +}+\xi_{y}(\partial_{5}W^{5+}-iM_{W}\phi^{+}\{1 + r_{\phi}\left( \delta(y) + \delta(y - \pi R)\right)\})\Big \vert ^2 ,\nonumber \\
\end{eqnarray}
In the above, $M_W$ is the $W$ boson mass and $\xi _y$ is related to {\em physical} gauge fixing parameter $\xi$ (taking values 1 in Feynman gauge and 0 in Landau gauge) via 
\begin{equation}
\xi= \xi_y \{1 + r_{\phi}\left( \delta(y) + \delta(y - \pi R)\right)\}.
\end{equation}

Before delving into the interactions needed for the calculation, let us spend some time discussing the 
physical eigenstates which are the outcome of mixing of some of the states originally present in the 
four dimensional effective theory. These are quite similar but not exactly the same as in UED. So we have decided to  make a dedicated discussion on this issue. There are two such cases relevant for our calculation. Let us first focus on the mixing in the quark sector. 
This mixing is driven by the Yukawa coupling thus it is only important and relevant for top quarks. 

Substituting  the modal expansions for fermions given in eq.\ref{fermionexpnsn}, in actions given in eq.\ref{actnquark} and eq.\ref{yukawa}  one can easily find the bilinear terms involving  the doublet and singlet states of the quarks. In $n$th KK-level, mass matrix 
reads  as 
\begin{equation}
\begin{pmatrix}
\bar{Q}_{t_{L}}^{(m)} & \bar{U}_{L}^{(m)}
\end{pmatrix}
\begin{pmatrix}
-M_{Qn}\delta^{mn} & m_{t} {\cal I}^{mn} \\ m_{t} {\cal I}^{mn}& M_{Qn}\delta^{mn}
\end{pmatrix}
\begin{pmatrix}
Q_{t_{R}}^{(n)} \\ U_{R}^{(n)}
\end{pmatrix}+{\rm h.c.},
\end{equation}
where $M_{Qn}$ are the solutions of transcendental equations as in eq.\ref{fermion_mass}. ${\cal I}^{mn}$ is an 
overlap integral of the form $$ \int_0 ^{\pi R}
\left[ 1+ r_y \delta(y) + r_y \delta(y - \pi R) \right] f_{L}^m (y) g_{R}^n (y)\;dy. $$
This integral is in general, non-zero for both $n=m$ and $n\neq m$. The second case would lead to the
(KK-)mode mixing among the quark of a particular flavour. However, the choice $r_y = r_f$  would 
make this integral equal to 1 (when $m =n$) or 0 ($m \neq n$). So by choosing equal fermion and Yukawa BLKTs one could easily avoid the mode mixing and end up in a simpler form of the fermion mixing matrix.  In the following we will stick to the choice of equal $r_y$ and $r_f$.

One can note that strength (off-diagonal terms) of the mixing is proportional to quark mass (denoted by $m_t$ here), hence  the mixing is only important for 
top quark (and we will denote top quark mass by $m_t$ in the following).  The resulting matrix can be diagonalized by separate unitary transformations for the left- and right-handed fields respectively:
\begin{equation}
\mathcal{U}_{L}^{(n)}=\begin{pmatrix}
-\cos\alpha_{n} & \sin\alpha_{n} \\ \sin\alpha_{n} & \cos\alpha_{n}
\end{pmatrix},~~\mathcal{U}_{R}^{(n)}=\begin{pmatrix}
\cos\alpha_{n} & -\sin\alpha_{n} \\ \sin\alpha_{n} & \cos\alpha_{n}
\end{pmatrix},
\end{equation}
where $\alpha_{n} = \frac12\tan^{-1}\left(\frac{m_{t}}{M_{Qn}}\right)$ is the mixing angle. Gauge eigenstates $Q_{t}^{(n)}$ and $U^{(n)}$ and mass eigenstates $Q_{t}^{\prime (n)}$ and $U^{\prime (n)}$ are related by,
\begin{eqnarray}
Q_{t_{L/R}}^{(n)} &=& \mp \cos\alpha_{n}Q_{t_{L/R}}^{\prime(n)}+\sin\alpha_{n}U_{L/R}^{\prime(n)},\\
U_{L/R}^{(n)} &=& \pm \sin\alpha_{n}Q_{t_{L/R}}^{\prime(n)}+\cos\alpha_{n}U_{L/R}^{\prime(n)},
\end{eqnarray}
where the mass eigenstates share the same mass eigenvalue $m_{Q_{t}^{\prime(n)}}=m_{U^{\prime(n)}}=\sqrt{m_{t}^{2}+M_{Qn}^{2}}=M_{u} \left(\rm say\right)$.

The four dimensional effective Lagrangian would also contain bilinear terms involving the KK-excitations (starting from KK-level $n =1$ and above) of the 5th components of $W^{\pm}$ ($Z$) bosons  and the KK-excitations of $\phi^\pm$ ($\chi^0$) of the Higgs doublet field \cite{petre}. 
In the following we note down the  bilinear terms involving the KK-modes of $W_5 ^{\pm n}
 $ and $\phi^{\pm n}$, which are relevant for our calculation. Using  eqs.\ref{gauge},\ref{higgs},  and eq.\ref{gauge_fix} one can write in the $R_{\xi}$ gauge,
 
 \begin{equation}
\mathcal{L}_{W_{5} ^{n \pm}  \phi^{n \mp} } = - \begin{pmatrix}
W_{5}^{(n)-} & \phi^{(n)-}
\end{pmatrix}
\begin{pmatrix}
M_{W}^{2}+\xi M_{\Phi n}^{2} & -i(1-\xi)M_{W}M_{\Phi n} \\ i(1-\xi)M_{W}M_{\Phi n} & M_{\Phi n}^{2}+\xi M_{W}^{2}
\end{pmatrix}
\begin{pmatrix}
W_{5}^{(n)+} \\ \phi^{(n)+}
\end{pmatrix}.
\end{equation}
The above mass matrix upon diagonalization would lead to a tower of 
charged Goldstone bosons (with mass square $\xi  (M_{\Phi n} ^2 + M_W^2) $),
 $$
G^{\pm (n)} = \frac{1}{M_{W_{n}}}\left(M_{\Phi n}W^{\pm5(n)}\mp  iM_{W}\phi^{\pm(n)}\right),
$$
and a physical charged Higgs pair (with mass square $M_{\Phi n} ^2 + M_W^2$):
$$
H^{\pm (n)} = \frac{1}{M_{W_{n}}}\left(M_{\Phi n}\phi^{\pm(n)}\mp iM_{W}W^{\pm5(n)}\right).
$$
So the fields $W^{\mu (n)\pm}$, $G^{(n)\pm}$ and $H^{(n)\pm}$ share the common mass eigenvalue $M_{W_{n}} \equiv
\sqrt {M_{\Phi n} ^2 + M_W^2}$ in 't-Hooft Feynman gauge ($\xi  = 1$).
These combinations of charged Higgs and Goldstone ensure the vanishing coupling of $\gamma H^{n \pm}W_{\nu}^{n \mp}$ as it should be with a doublet Higgs at our dispense.

Necessary interactions involving the $Z$-boson, fermions and scalars in the four dimensional effective theory can be derived from the above action by simply 
inserting the appropriate $y$ dependent profile for the respective five dimensional fields and then integrating over the extra direction, $y$. In contrast to mUED, where $y$ dependent profiles are either $\sin (\frac{ ny}{R}) $ or $\cos (\frac{ ny}{R})$, some of the couplings in nmUED are hallmarked by the presence of few overlap integrals of the form :
\begin{equation}
I^{mn} = \int_0 ^{\pi R}
dy \;f_{\alpha}^n (y) f_{\beta}^m  (y) f_{\rho}^p (y)
\end{equation}
Here, greek indices refer to the kind of fields involved in the coupling while roman indices refer to the KK-level of respective fields. 

At this end, let us pay some attention to a pair of overlap integrals $I ^{mn} _1$ and $I^{mn} _2$ which are relevant for our calculation appearing in the interactions listed in appendix A. $I^{mn}_{1}$ and $I^{mn}_{2}$ are the following overlap integrals:  
\begin{eqnarray}
I ^{mn}_{1} &=& \int_{0}^{\pi R} dy \;[1+r_{f}\{\delta(y)+\delta(y-\pi R)\}]f_{Q_{t_{L}}}^{(m)} f_{\phi}^{(n)}f_{b_{L}}^{(0)},\\
I ^{mn}_{2} &=& \int_{0}^{\pi R} dy~f_{Q_{t_{R}}}^{(m)} f_{W_{5}}^{(n)}f_{b_{L}}^{(0)}. 
\end{eqnarray}

These integrals are non-zero when $n+m$ is even. Integrals and interactions among KK-states with odd $n + m$ identically vanish due to a conserved KK-parity. Even in the former case, the integrals are of the order 1 when $n=m$.  When $m$ differs from $n$ (in the case of even $n+m$) values of the integrals diminish generally by an order of magnitude than the $m=n$ case\footnote{As for example, when $r_f = 1$ and $r_{\phi} = 2$: $I_{1}^{11} =0.82$, $I_{1}^{22} =0.88$, $I_{1}^{33} =0.92$, $I_{1}^{44} =0.94$, $I_{1}^{55} =0.96$, $I_{1}^{31} =0.01$, $I_{1}^{51} =0.004$, $I_{1}^{53} =0.03$, $I_{1}^{42} =0.03$; $I_{2}^{11} \sim I_{2}^{22} \sim I_{2}^{33} \sim I_{2}^{44} \sim I_{2}^{55} = 0.99$ and $I_{2}^{31} = 0.07$, $I_{2}^{51} = 0.02$, $I_{2}^{42} = 0.08$.}. Keeping this in mind we will be only considering the interactions with $n=m$ neglecting the other sub dominant contributions coming from interactions in which $n\neq m$. The expressions for the integrals (upon integrating over $y$) are given in appendix A along with the necessary Feynman rules.

\section{Calculation of Radiative Correction to the $Zb \bar b$ vertex:}
We are now all set to discuss the detail of the calculation leading to the correction of the $Zb\bar b$ vertex in the framework of nmUED.  
However, as a preamble we will first briefly discuss 
the meaning of $R_b$ and its correlation to $Zb\bar{b}$ coupling in the SM. The tree level $Zb\bar{b}$ coupling, in the SM, can be defined as
\begin{equation}
\frac{g}{\cos\theta_{W}}\bar{b}^{0}\gamma^{\mu}(g^0 _{L}P_{L}+ g^0 _{R}P_{R})b^{0}Z_{\mu}^{0},
\end{equation}
where $Z_{\mu}^{0}$ and $b^{0}$'s are SM fields, $P_{R,L}=(1 \pm \gamma_{5})/2$ are the right- and left-chirality projectors respectively and
\begin{eqnarray}
g^0 _{L} &=& -\frac{1}{2}+\frac{1}{3}\sin^{2}\theta_{W},\\
g^0 _{R} &=& \frac{1}{3}\sin^{2}\theta_{W}.
\end{eqnarray}
Any higher order quantum corrections either from SM or from new physics (NP) can be incorporated uniformly as the modification to this tree level couplings 
 given as
\begin{eqnarray}
g_{L} &=&  g^0 _L + \delta g_{L}^{\rm SM} + \delta g_{L}^{\rm NP},\\
g_{R} &=&  g^0 _R + \delta g_{R}^{\rm SM} + \delta g_{R}^{\rm NP},
\end{eqnarray}
where $\delta g_{L/R}^{\rm SM}$ are the radiative corrections from SM and $\delta g_{L/R}^{\rm NP}$ are that of NP \cite{oliver_zbb}.
These corrections can modify the $Z$ decay width to $b$ quarks normalized to the total hadronic decay width of $Z$, defined by a dimensionless variable, 
\begin{equation}
R_{b} \equiv \frac{\Gamma(Z \to b\bar{b})}{\Gamma(Z \to {\rm hadrons})}.
\end{equation}
We will only be considering the effect due to  the third generation quarks. Normally, at the one loop order (SM \& also in NPs) only the $g_{L}$ receives correction proportional to $m_{t}^{2}$, and the $g_{R}$ receives correction proportional to $m_{b}^{2}$ (due to the difference in couplings between two chiralities) where $m_{t}$ ($m_{b}$) is the zero mode top (bottom) quark mass. We have neglected the $b$ mass in our calculation and thus a shift $\delta g_{L}^{\rm NP}$ translates into a shift in $R_{b}$ given by,
\begin{equation}
\delta R_{b} = 2R_{b}(1-R_{b})\frac{\hat g_{L}}{\hat g_{L}^{2}+\hat g_{R}^{2}}\delta g_{L}^{\rm NP},
\label{rb}
\end{equation}
with $\hat g_L$ and $\hat g_R$ given by 
\begin{eqnarray}
\hat g_{L}^{b} &=& \sqrt{\rho_b}\,(-\frac{1}{2}+\kappa_{b}\,\frac{1}{3}\,\sin^{2}\theta_{W}), \nonumber \\
\hat g_{R}^{b} &=& \frac{1}{3}\sqrt{\rho_b}\,\kappa_{b}\,\sin^{2}\theta_{W}, \nonumber
\end{eqnarray}
after incorporating the SM electroweak corrections   only \cite{Beringer}.  Here, $\rho_{b} = 0.9869$ and 
$\kappa_{b} = 1.0067$.

In general, the $g_{L}^{\rm NP}$ is calculable in a given framework while 
$R_b$ is an experimentally measurable quantity. Thus eq.\ref{rb} can be  used to constrain the parameters of the model.  We would exactly like to do this exercise in the framework of nmUED in the following.

\begin{figure}[thb]

 \centering
 \subfigure[]{
   \includegraphics[scale=0.5]{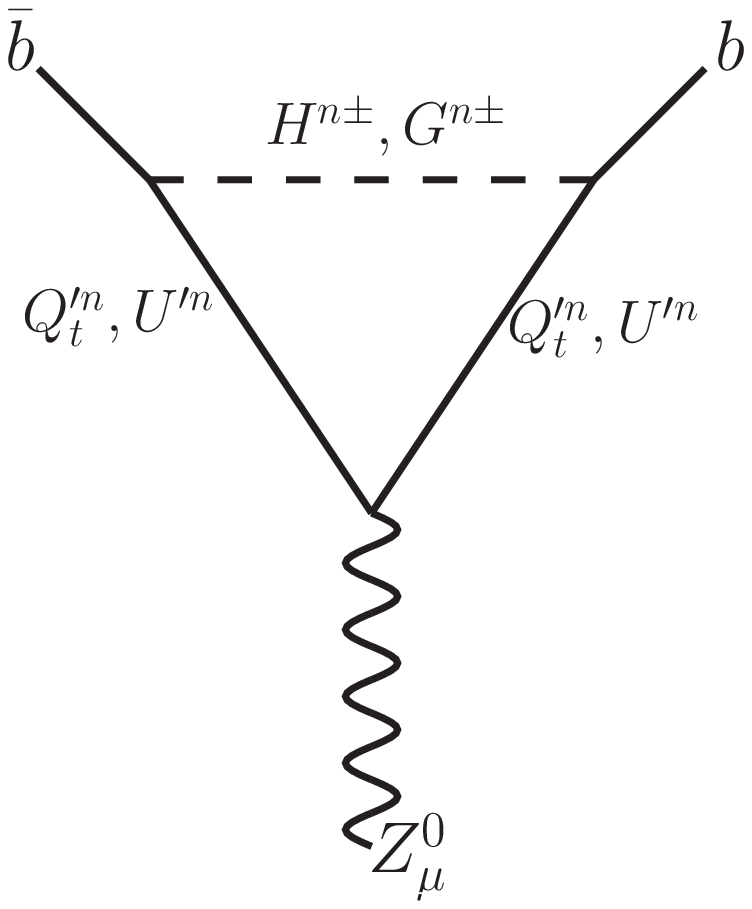}    }
     \subfigure[]{
   \includegraphics[scale=0.5]{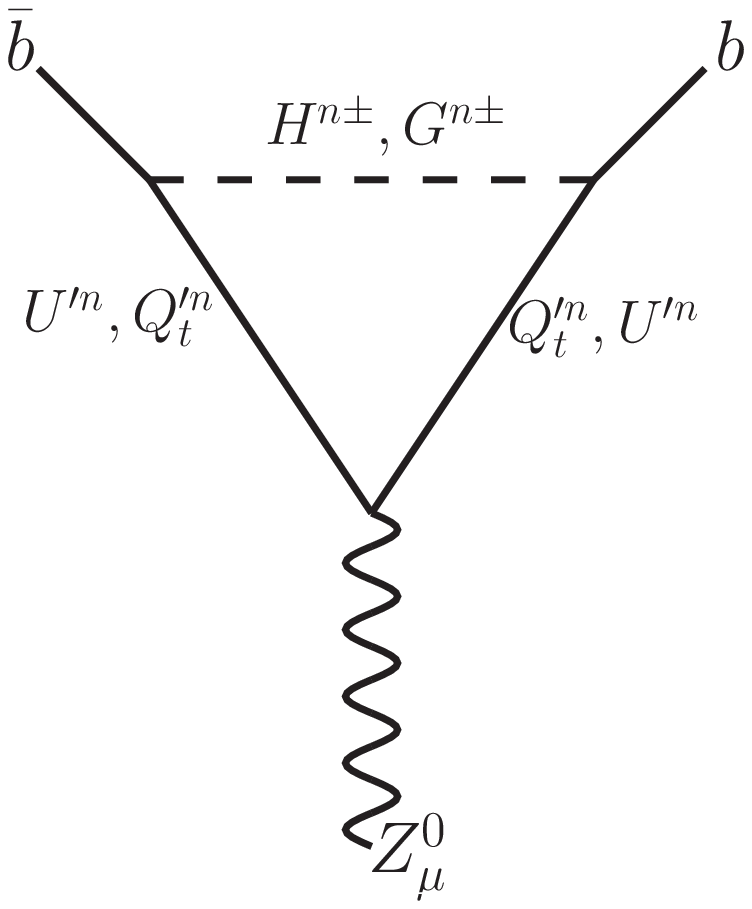}
   }
  \subfigure[]{
   \includegraphics[scale=0.5]{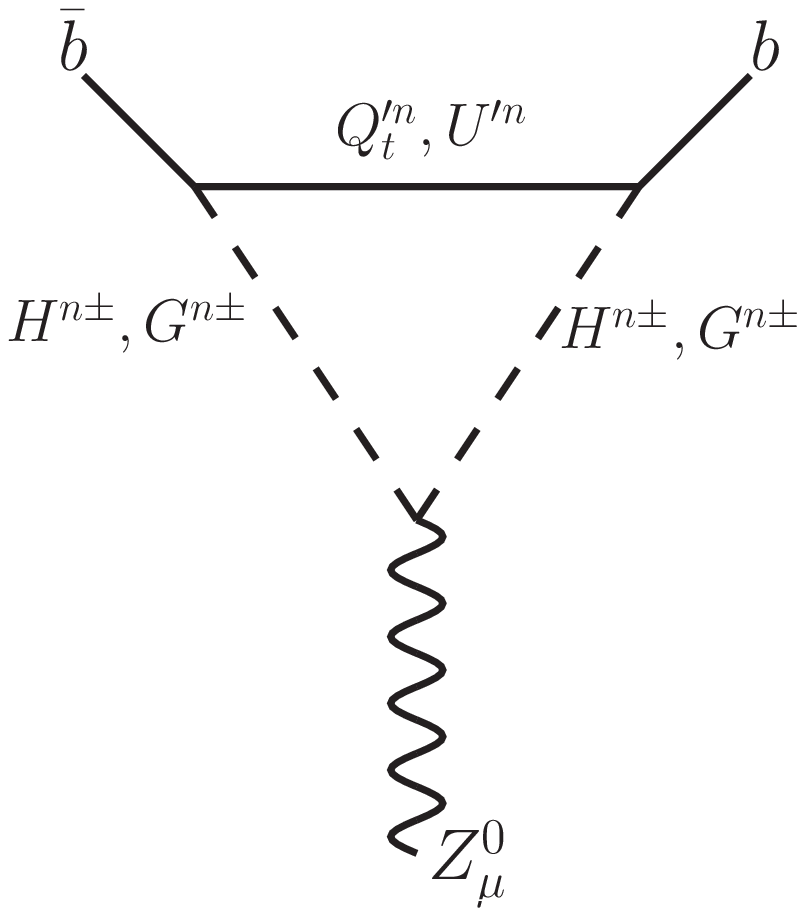}
   } \\
  \subfigure[]{
   \includegraphics[scale=0.5]{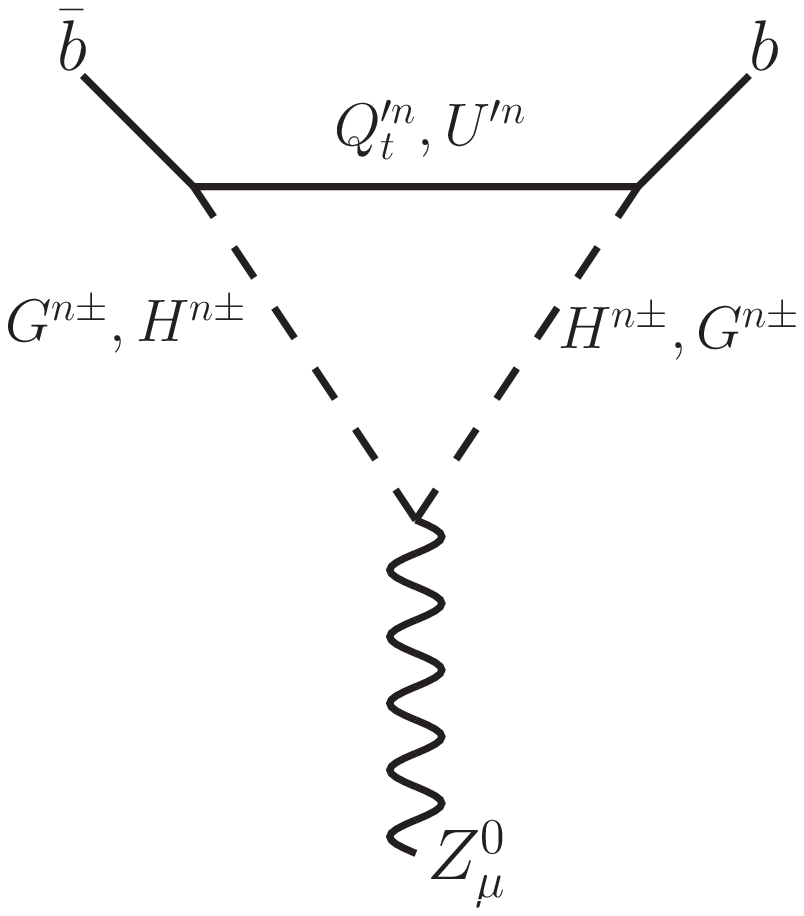}
    }
 \subfigure[]{
   \includegraphics[scale=0.5]{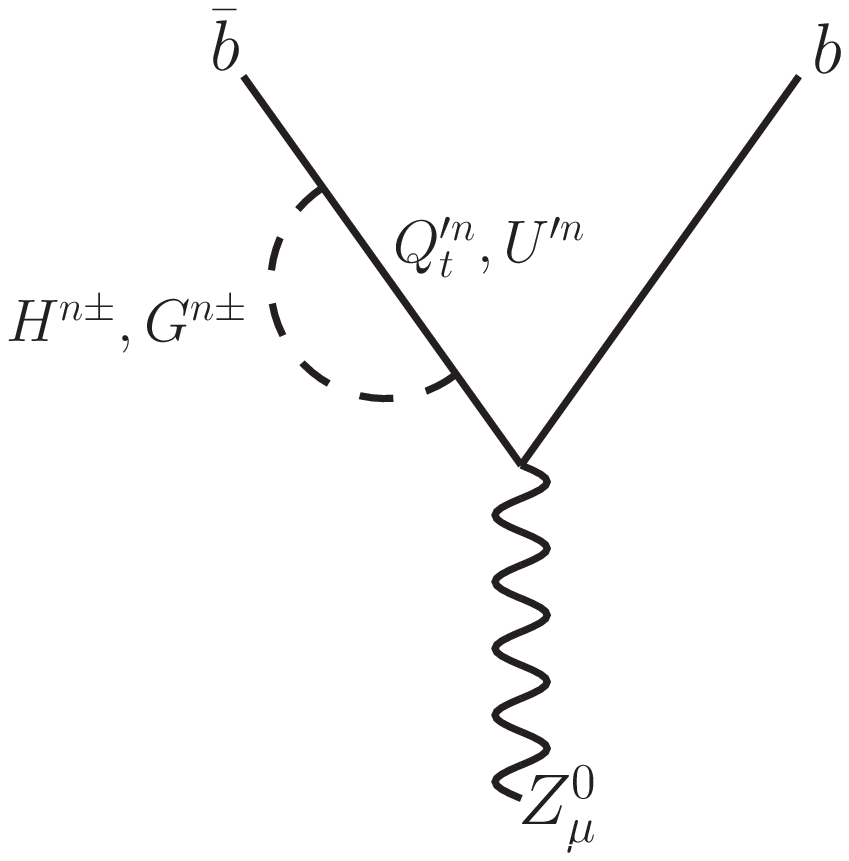}
   }
     \subfigure[]{
    \includegraphics[scale=0.5]{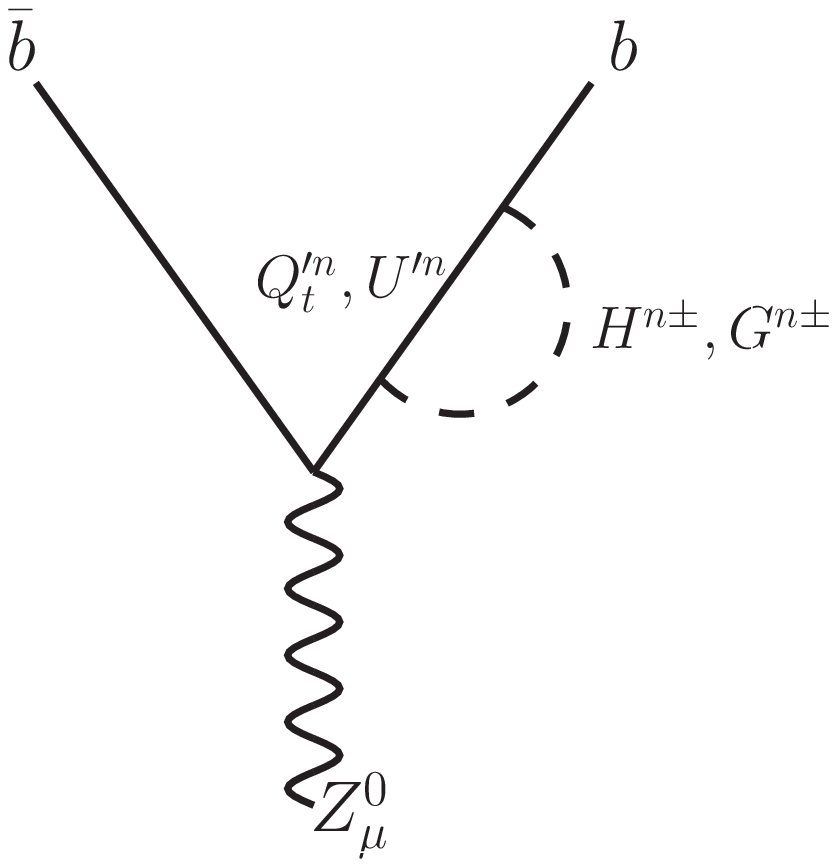}
   }
    \caption[]{Loop involving KK-mode of scalar and fermion propagators.}
    \label{fig1}
\end{figure}

Since we have neglected the interactions involving KK-states with unequal KK-numbers in an interaction vertex, the number of diagrams contributing to radiative corrections of the $Zb\bar{b}$ vertex in nmUED are same as that of minimal UED. Fig.\ref{fig1} shows the Feynman diagrams involving KK-excitations of top quarks, charged Higgs/Goldstone bosons in the loop. The contribution coming from the diagrams of Fig.\ref{fig1} is dominant for the presence of Yukawa coupling which is proportional to $m_t$. In our calculations, we have considered momentum of each external leg to be zero and have neglected the $b$ quark mass. The amplitude of each diagram, for $n$th KK-mode, can be expressed in terms of a single function, $f^{n}(r_{n},r_{n}^{\prime},M^{\prime})$, defined as,
\begin{equation}
i {\mathcal M}^{(n)}=
i \frac{g}{\cos\theta_{W}} \overline{u}(p_{1},s_{1})  f^{n}(r_{n},r_{n}^{\prime},M^{\prime}) \gamma^\mu P_L v(p_{2},s_{2}) \epsilon_\mu(q)~,
\end{equation}
where $r_{n} \equiv m_t^2/M_{Qn}^2$ , $r_{n}^{\prime} \equiv M_W^2/M_{Qn}^2$ , $M^{\prime} \equiv M_{\Phi n}^{2}/M_{Qn}^2$.
 
Amplitudes of different diagrams of Fig.\ref{fig1} (evaluated in 't-Hooft- Feynman gauge) are given by,
\begin{eqnarray}
f _{1(a)}^{n}(r_{n},r_{n}^{\prime},M^{\prime}) &=& \frac{\beta}{(4\pi)^2}  \frac{g^{2}}{8}\{-\frac{4}{3}\sin^{2}\theta_{W}\left(I_{2}^{2}+ I_{1}^{2}\frac{m_{t}^{2}}{M_{W}^{2}}\right) + I_{2}^{2}\left(\cos^{4}\alpha_{n} + \sin^{4}\alpha_{n}\right) \nonumber \\
& & + 2I_{1}^{2} \frac{m_{t}^{2}}{M_{W}^{2}}\sin^{2}\alpha_{n}\cos^{2}\alpha_{n} \}\Big[\delta_{n} -1 +
\{5(r_{n}+1)^{2} + 3(r_{n}^{\prime}+M^{\prime})^{2} \nonumber \\
& & - 8(r_{n}+1)(r_{n}^{\prime}+M^{\prime}) - 2{(1+r_{n})}^{2}\ln(1+r_{n}) \nonumber \\
 & & - 2{(M^{\prime}+r_{n}^{\prime})}^{2}\ln(M^{\prime}+r_{n}^{\prime}) + 4(1 + r_{n})(M^{\prime} + r_{n}^{\prime})\ln(M^{\prime}+r_{n}^{\prime})\} \nonumber \\
 & & /2\{\left(r_{n}+1\right)-\left(M^{\prime}+r_{n}^{\prime}\right)\}^{2}\Big],\\
f_{1(b)}^{n}(r_{n},r_{n}^{\prime},M^{\prime})  &=& \frac{\beta}{(4\pi)^2}  \frac{g^{2}}{8}\{2I_{2}^{2}\sin^{2}\alpha_{n}\cos^{2}\alpha_{n}
- 2I_{1}^{2}\frac{m_{t}^{2}}{M_{W}^{2}}\sin^{2}\alpha_{n}\cos^{2}\alpha_{n} \} \nonumber \\
 & & \Big[\delta_{n} -1 + \{-3(r_{n}+1)^{2} + 3(r_{n}^{\prime}+M^{\prime})^{2}\nonumber \\
 & & - 2{(1+r_{n})}^{2}\ln(1+r_{n}) - 2{(M^{\prime}+r_{n}^{\prime})}^{2}\ln(M^{\prime}+r_{n}^{\prime}) \nonumber \\
 & &  + 8(1 + r_{n})(M^{\prime} + r_{n}^{\prime})\ln(1+r_{n}) \nonumber \\
 & & - 4(1 + r_{n})(M^{\prime} + r_{n}^{\prime})\ln(M^{\prime}+r_{n}^{\prime})\}/2\{\left(r_{n}+1\right)-\left(M^{\prime}+r_{n}^{\prime}\right)\}^{2}\Big],\\
f_{1(c+d)}^{n}(r_{n},r_{n}^{\prime},M^{\prime})  &=& \frac{\beta}{(4\pi)^2}  \frac{g^{2}}{8}\{\left(-1+2\sin^{2}\theta_{W}\right)\left(I_{2}^{2}+ I_{1}^{2}\frac{m_{t}^{2}}{M_{W}^{2}}\right) - I_{2}^{2}\} \nonumber \\
 & & \Big[\delta_{n} + \{3(r_{n}+1)^{2} + (r_{n}^{\prime}+M^{\prime})^{2} \nonumber \\
& & - 4(r_{n}+1)(r_{n}^{\prime}+M^{\prime}) - 2{(1+r_{n})}^{2}\ln(1+r_{n}) \nonumber \\
 & & - 2{(M^{\prime}+r_{n}^{\prime})}^{2}\ln(M^{\prime}+r_{n}^{\prime}) + 4(1 + r_{n})(M^{\prime} + r_{n}^{\prime})\ln(M^{\prime}+r_{n}^{\prime})\} \nonumber \\
 & & /2\{\left(r_{n}+1\right)-\left(M^{\prime}+r_{n}^{\prime}\right)\}^{2}\Big],\\
f_{1(e+f)}^{n}(r_{n},r_{n}^{\prime},M^{\prime}) &=& \frac{\beta}{(4\pi)^2}  \frac{g^{2}}{8} \left(1-\frac{2}{3}\sin^{2}\theta_{W}\right)\left(I_{2}^{2}+ I_{1}^{2}\frac{m_{t}^{2}}{M_{W}^{2}}\right) \nonumber \\
  & & \Big[\delta_{n} + \{3(r_{n}+1)^{2} + (r_{n}^{\prime}+M^{\prime})^{2} \nonumber \\
& & - 4(r_{n}+1)(r_{n}^{\prime}+M^{\prime}) - 2{(1+r_{n})}^{2}\ln(1+r_{n}) \nonumber \\
 & & - 2{(M^{\prime}+r_{n}^{\prime})}^{2}\ln(M^{\prime}+r_{n}^{\prime}) + 4(1 + r_{n})(M^{\prime} + r_{n}^{\prime})\ln(M^{\prime}+r_{n}^{\prime})\} \nonumber \\
 & & /2\{\left(r_{n}+1\right)-\left(M^{\prime}+r_{n}^{\prime}\right)\}^{2}\Big],
\end{eqnarray}
where $\delta_{n} \equiv 2/\epsilon - \gamma + \log(4\pi) + \log(\mu^2/M_{Qn}^2),$ and $\mu$ is the 't-Hooft mass scale; $\beta \equiv \frac{\pi R + r_\phi}{\pi R + r_f}$. $I_{1}$ and $I_{2}$ stand for the overlap integrals given in equations $(20)$ and $(21)$ respectively for $n=m$. Amplitudes of diagrams ($e$) and ($f$) are multiplied by a factor of $\frac{1}{2}$ which  comes from the usual convention of contributing one-half of this correction into self-energy and the other half in the wave function renormalization. Total amplitude ($i\mathcal M_{1}^{(n)}$) of  diagrams in Fig.\ref{fig1} is obtained by adding the individual amplitudes for each diagram and is given by the following expression:
\begin{eqnarray}
i {\mathcal M}_{1}^{(n)} &=& \frac{i}{(4\pi)^2}  \frac{g^{3}}{4\cos\theta_{W}} \overline{u}\left(p_{1},s_{1}\right)\frac{r_{n}\beta}{\{\left(r_{n}+1\right)-\left(M^{\prime}+r_{n}^{\prime}\right)\}^{2}}\left(-I_{2}^{2} + I_{1}^{2}\frac{m_{t}^{2}}{M_{W}^{2}}\right)\nonumber \\
 & & \Big[(1+r_{n}) - (M^{\prime}+r_{n}^{\prime}) + (M^{\prime} + r_{n}^{\prime})\ln\left(\frac{M^{\prime}+r_{n}^{\prime}}{1+r_{n}}\right)\Big]\gamma^\mu P_L v\left(p_{2},s_{2}\right) \epsilon_{\mu}(q).
\end{eqnarray}

From the above expression, it is evident that terms proportional to $\delta_{n}$, as well as $\sin^{2}\theta_{W}$ cancel among themselves. Here, any correction proportional to $\sin ^2 \theta_{W}$ in the $Zb\bar{b}$ coupling must be reflected in the renormalization of charge (of $b$ quark). This implies that any finite renormalization to  the $\gamma b\bar{b}$ vertex must be the same to any correction proportional to $ \sin ^2 \theta_{W}$ in the $Zb\bar{b}$ vertex. We have explicitly checked that both of these corrections coming from diagrams of the same topology depicted in Fig.\ref{fig1} identically vanishes. 

\begin{figure}[h]
 \centering
 \subfigure[]{
   \includegraphics[scale=0.5]{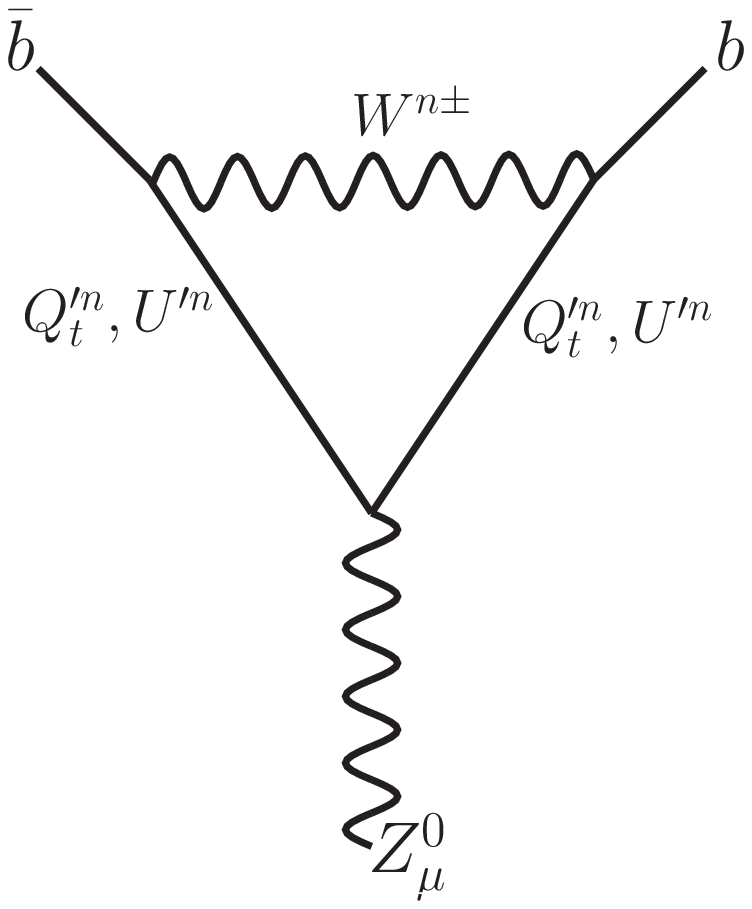}    }
     \subfigure[]{
   \includegraphics[scale=0.5]{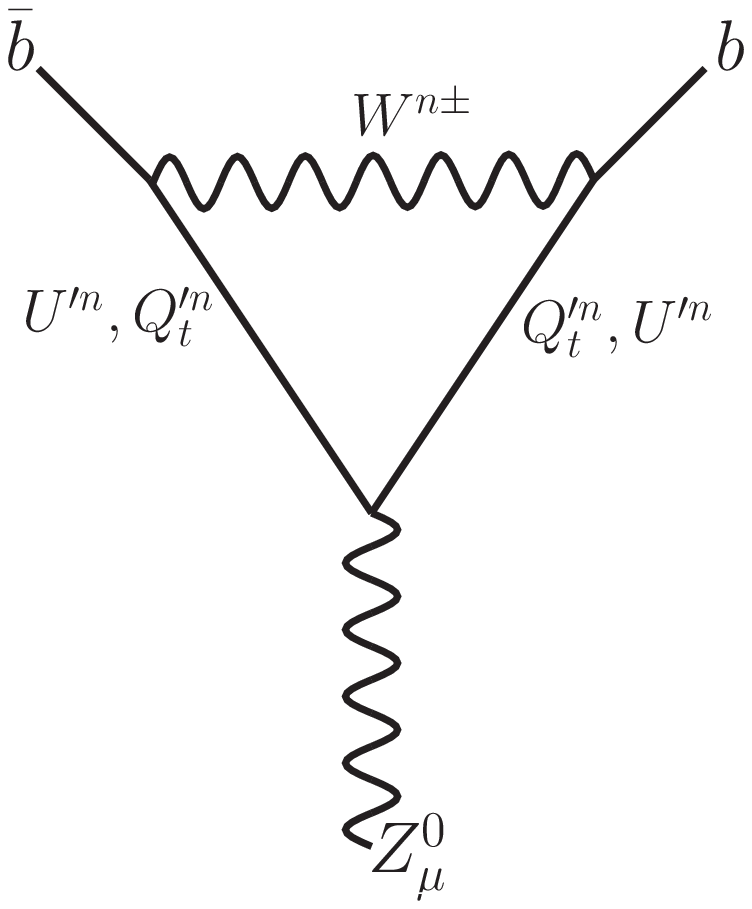}
   }
  \subfigure[]{
   \includegraphics[scale=0.5]{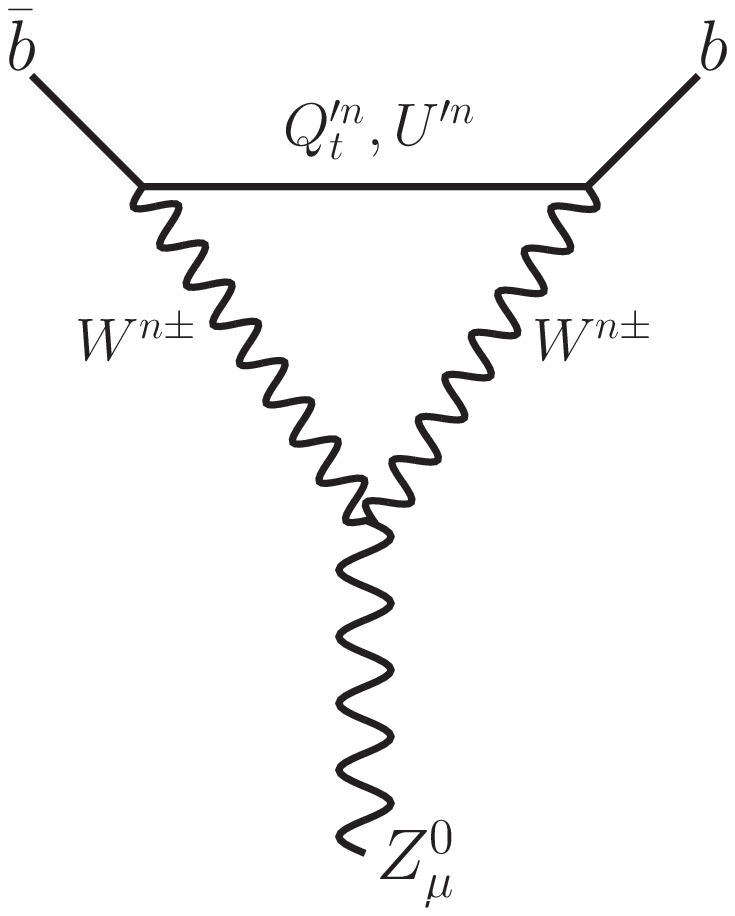}
   } \\
  \subfigure[]{
   \includegraphics[scale=0.5]{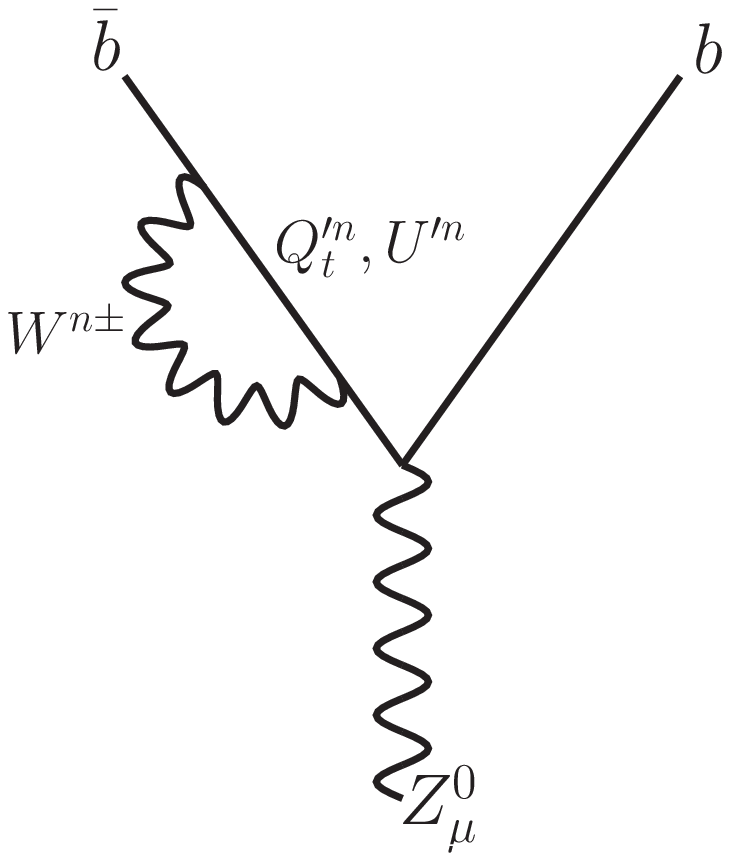}
    }
 \subfigure[]{
   \includegraphics[scale=0.5]{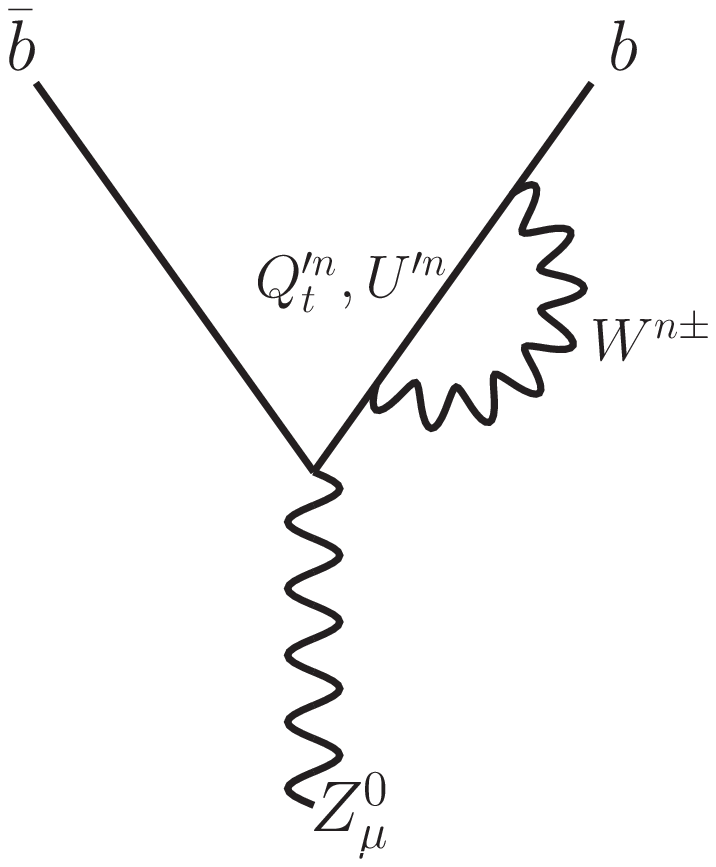}
   }
     \subfigure[]{
    \includegraphics[scale=0.5]{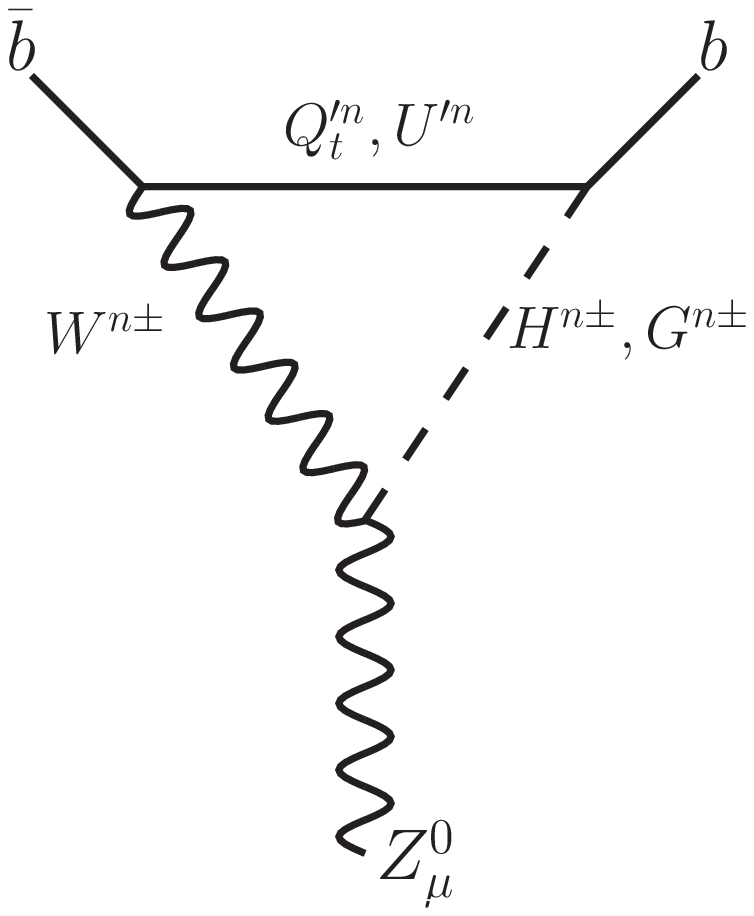}
   }
    \caption[]{Loop involving KK-mode of W and Goldstone propagators.}
    \label{fig2}
\end{figure}

There is a second set of diagrams contributing to effective $Zb\bar{b}$ interaction mainly arising from the KK-excitations of $W$ 
bosons and quarks. These are sub dominant with respect to the contributions coming from Fig.\ref{fig1}.

 In the following we present the amplitudes  of all the individual diagrams  given in Fig.\ref{fig2} :
\begin{eqnarray}
f_{2(a)}^{n}(r_{n},r_{n}^{\prime},M^{\prime}) &=& \frac{I_{1}^{2}\beta}{(4\pi)^2}  \frac{g^{2}}{4}\{-\frac{4}{3}\sin^{2}\theta_{W} + \cos^{4}\alpha_{n} + \sin^{4}\alpha_{n} \} \Big[\delta_{n} -2 \nonumber \\
 & & +
\{5(r_{n}+1)^{2} + 3(r_{n}^{\prime}+M^{\prime})^{2} \nonumber \\
& & - 8(r_{n}+1)(r_{n}^{\prime}+M^{\prime}) - 2{(1+r_{n})}^{2}\ln(1+r_{n}) \nonumber \\
 & & - 2{(M^{\prime}+r_{n}^{\prime})}^{2}\ln(M^{\prime}+r_{n}^{\prime}) + 4(1 + r_{n})(M^{\prime} + r_{n}^{\prime})\ln(M^{\prime}+r_{n}^{\prime})\} \nonumber \\
 & & /2\{\left(r_{n}+1\right)-\left(M^{\prime}+r_{n}^{\prime}\right)\}^{2}\Big], \\
f_{2(b)}^{n}(r_{n},r_{n}^{\prime},M^{\prime}) &=& \frac{I_{1}^{2}\beta}{(4\pi)^2}  \frac{g^{2}}{4}\{2\sin^{2}\alpha_{n}\cos^{2}\alpha_{n}\} \nonumber \\
 & & \Big[\delta_{n} -2 + \{-3(r_{n}+1)^{2} + 3(r_{n}^{\prime}+M^{\prime})^{2}\nonumber \\
 & & - 2{(1+r_{n})}^{2}\ln(1+r_{n}) - 2{(M^{\prime}+r_{n}^{\prime})}^{2}\ln(M^{\prime}+r_{n}^{\prime}) \nonumber \\
 & &  + 8(1 + r_{n})(M^{\prime} + r_{n}^{\prime})\ln(1+r_{n}) - 4(1 + r_{n})(M^{\prime} + r_{n}^{\prime})\ln(M^{\prime}+r_{n}^{\prime})\} \nonumber \\
 & & /2\{\left(r_{n}+1\right)-\left(M^{\prime}+r_{n}^{\prime}\right)\}^{2}\Big], \\
f_{2(c)}^{n}(r_{n},r_{n}^{\prime},M^{\prime}) &=& -\frac{I_{1}^{2}\beta}{(4\pi)^2}  \frac{g^{2}}{4}\left(6\cos^{2}\theta_{W}\right) \Big[\delta_{n} -\frac{2}{3} + \{3(r_{n}+1)^{2} + (r_{n}^{\prime}+M^{\prime})^{2} \nonumber \\
& & - 4(r_{n}+1)(r_{n}^{\prime}+M^{\prime}) - 2{(1+r_{n})}^{2}\ln(1+r_{n}) \nonumber \\
 & & - 2{(M^{\prime}+r_{n}^{\prime})}^{2}\ln(M^{\prime}+r_{n}^{\prime}) + 4(1 + r_{n})(M^{\prime} + r_{n}^{\prime})\ln(M^{\prime}+r_{n}^{\prime})\} \nonumber \\
 & & /2\{\left(r_{n}+1\right)-\left(M^{\prime}+r_{n}^{\prime}\right)\}^{2}\Big],\\
f_{2(d+e)}^{n}(r_{n},r_{n}^{\prime},M^{\prime}) &=& \frac{I_{1}^{2}\beta}{(4\pi)^2}  \frac{g^{2}}{4}\left(1-\frac{2}{3}\sin^{2}\theta_{W}\right)\Big[\delta_{n} -1 + \{3(r_{n}+1)^{2} + (r_{n}^{\prime}+M^{\prime})^{2} \nonumber \\
& & - 4(r_{n}+1)(r_{n}^{\prime}+M^{\prime}) - 2{(1+r_{n})}^{2}\ln(1+r_{n}) \nonumber \\
 & & - 2{(M^{\prime}+r_{n}^{\prime})}^{2}\ln(M^{\prime}+r_{n}^{\prime}) + 4(1 + r_{n})(M^{\prime} + r_{n}^{\prime})\ln(M^{\prime}+r_{n}^{\prime})\} \nonumber \\
 & & /2\{\left(r_{n}+1\right)-\left(M^{\prime}+r_{n}^{\prime}\right)\}^{2}\Big],\\
f_{2(f)}^{n}(r_{n},r_{n}^{\prime},M^{\prime}) &=& \frac{I_{1}^{2}\beta}{(4\pi)^2}g^{2}\{(r_{n} +1)\sin^{2}\theta_{W} -1\}\nonumber \\
 & & \{-(1+r_{n}) + (M^{\prime}+r_{n}^{\prime}) + (1 + r_{n})\ln\left(\frac{1+r_{n}}{M^{\prime}+r_{n}^{\prime}}\right)\}\nonumber \\
& & /\{\left(r_{n}+1\right)-\left(M^{\prime}+r_{n}^{\prime}\right)\}^{2}\Big].
\end{eqnarray}

 In diagrams of Fig.\ref{fig2}, divergences along with the terms proportional to $\sin^{2}\theta_{W}$ do not cancel among themselves. The divergent  terms are $r_{n}$ independent.  Following the prescription in Ref.  \cite{pichsanta}, we can write the renormalized amplitude as:
\begin{equation}
i {\mathcal M}_{2R}^{(n)}(r_{n},r_{n}^{\prime},M^{\prime}) = i {\mathcal M}_{2}^{(n)}(r_{n},r_{n}^{\prime},M^{\prime})- i {\mathcal M}_{2}^{(n)}(r_{n}=0,r_{n}^{\prime},M^{\prime}).
\end{equation} 
Finally summing up contributions  from all diagrams we have,
\begin{eqnarray}
& & i {\mathcal M}_{\rm total}^{(n)} = i {\mathcal M}_{1}^{(n)} + i {\mathcal M}_{2R}^{(n)} = \frac{i}{(4\pi)^2}  \frac{g^{3}}{4\cos\theta_{W}} \overline{u}\left(p_{1},s_{1}\right)\frac{r_{n}\beta}{\{\left(r_{n}+1\right)-\left(M^{\prime}+r_{n}^{\prime}\right)\}^{2}} \nonumber \\ 
& & \Bigg[\left\{-I_{2}^{2} + I_{1}^{2}\left(-2+\frac{m_{t}^{2}}{M_{W}^{2}}\right)\right\}
\left\{(1+r_{n})-(r_{n}^{\prime}+M^{\prime})+(r_{n}^{\prime}+M^{\prime})\ln\left(\frac{r_{n}^{\prime}+M^{\prime}}{1+r_{n}}\right)\right\} \nonumber \\
 & & +4I_{1}^{2}\left\{-(1+r_{n})+(r_{n}^{\prime}+M^{\prime})+(1+r_{n})
\ln\left(\frac{1+r_{n}}{r_{n}^{\prime}+M^{\prime}}\right)\right\}\Bigg]\nonumber \\
 & & \gamma^\mu P_L v\left(p_{2},s_{2}\right) \epsilon_{\mu}(q). 
\end{eqnarray}
Therefore for each mode, correction in $g_{L}$:
\begin{eqnarray}
& & \delta g_{L}^{(n)\rm NP} = \Sigma _{i}f_{i}^{n}(r_{n},r_{n}^{\prime},M^{\prime}) =\frac{\sqrt{2}G_{F}m_{t}^{2}}{16\pi^{2}} F_{\rm nmUED}^{(n)}(r_{n},r_{n}^{\prime},M^{\prime}),
\label{gl}
\end{eqnarray}
where 
\begin{eqnarray}
F_{\rm nmUED}^{(n)} (r_{n},r_{n}^{\prime},M^{\prime}) &=& \frac{r_{n}\beta}{[(1+r_{n})-(r_{n}^{\prime}+M^{\prime})]^{2}}\Bigg[\left\{I_{1}^{2}\left(
1-\frac{2M_{W}^{2}}{m_{t}^{2}}\right)-I_{2}^{2}\frac{M_{W}^{2}}{m_{t}^{2}}\right\}\nonumber \\
& & \times \left\{(1+r_{n})-(r_{n}^{\prime}+M^{\prime})+(r_{n}^{\prime}+M^{\prime})\ln\left(\frac{r_{n}^{\prime}+m^{\prime}}{1+r_{n}}\right)\right\}\nonumber \\
& & +\frac{4M_{W}^{2}}{m_{t}^{2}}I_{1}^{2}\left\{-(1+r_{n})+(r_{n}^{\prime}+M^{\prime})+(1+r_{n})\ln\left(\frac{1+r_{n}}{r_{n}^{\prime}+M^{\prime}}\right)\right\}\Bigg].
\label{fued}
\end{eqnarray}
Total new physics contribution $\delta g_L ^{\rm NP}$ (and similarly $F_{\rm nmUED}$) can be obtained by summing $\delta g_{L}^{(n)\rm NP}$ over 
KK-modes ($n$).  It can be checked that the new physics contribution $\delta g_{L}^{\rm NP}$ and hence $F_{\rm nmUED}$ goes to zero when $R^{-1} \to \infty$, as expected in a decoupling theory. 
\section{Results}
Let us begin the discussion of our results with the present status of experimental and theoretical estimation of the $Zb\bar{b}$ coupling. 
Following Gfitter Collaboration \cite{Rb-Gfitter} and an improved estimation \cite{frietas} of $R_b$ incorporating higher order effects in the framework of SM, the experimental and the theoretical (SM)
values are $$ R_b ^{\rm exp}= 0.21629 \pm 0.00066  \;\;\; {\rm and}\;\;\;  R_b ^{\rm SM}= 0.21550 \pm 0.00003 .$$
Above results imply an 1.2 {\em standard deviation}  discrepancy between the experimental value of $R_b$ and its SM estimate. 
Thus, Eq.\ref{gl}, \ref{fued} in conjunction with Eq.\ref{rb} could
be used to translate this 1.2$\sigma$ discrepancy on $R_b$ to an allowed range for $F_{\rm nmUED}$: $-0.3165 \pm 0.2647$.   
One can now easily use this to constrain the model parameters of nmUED. 

Dominant contribution to $F_{\rm nmUED}$  comes from Feynman graphs listed in Fig.\ref{fig1}; as all of the 
amplitudes listed in Fig.\ref{fig1} contain terms proportional to $ g y_t ^2$.   While the contributions from  diagrams in Fig.\ref{fig2} are 
proportional only to $g^3$, with an exception to the diagram $2(f)$; which has some terms proportional to $ g^2 y_t$ (here, $y_t$ is the top quark Yukawa
coupling).  Total contribution of diagrams in Fig.\ref{fig1} is nearly $1.5$ times of the total contribution of the diagrams in Fig.\ref{fig2}.
 
\subsection{Relook at the bound on $R^{-1}$ in mUED from $R_b$}
Before we present our main results in the framework of nmUED, we would like to give a look at the limit on the $R^{-1}$, in case of
 UED, keeping in mind the  new estimate of 
SM radiative corrections to the $Zb\bar{b}$ vertex at two loop level \cite{frietas}. One can easily retrieve the UED contributions to $\delta g^{\rm NP}_L$ by simply setting BLKT parameters to zero. In this limit, overlap integrals ($I_1$ and $I_2$) used in the couplings become 
unity and $M_{Qn}, M_{gn}$ and $M_{\Phi n}$
all become equal to $n\over R$ in the $n$th KK-level;  the ratios $\beta$, $M^{\prime}$ will be unity and our expressions completely agree with those given in Ref.\cite{buras}. One can define a function $F^{(n)}_{\rm UED}$ in the same spirit following Eq.\ref{fued}:
\begin{eqnarray}
F_{\rm UED}^{(n)}(r_{1n},r_{1n}^{\prime})= & & \frac{r_{1n}}{(r_{1n}-r_{1n}^{\prime})^{2}}
\Bigg[\left(1-3\frac{M_{W}^{2}}{m_{t}^{2}}\right)\{(r_{1n}-r_{1n}^{\prime})+(1+r_{1n}^{\prime})\ln\left(\frac{1+r_{1n}^{\prime}}{1+r_{1n}}\right)\}\nonumber \\
& & +\frac{4M_{W}^{2}}{m_{t}^{2}}\{r_{1n}^{\prime}-r_{1n}+(1+r_{1n})\ln \left(\frac{1+r_{1n}}{1+r_{1n}^{\prime}}\right)\}\Bigg].
\label{fuedeq}
\end{eqnarray}
Here, $r_{1n} \equiv m_t^2/m_{n}^2$ , $r_{1n}^{\prime} \equiv M_W^2/m_{n}^2$ and $m_{n} = \frac{n}{R}$.

\begin{figure}[h]

 \centering
   \includegraphics[scale=0.4]{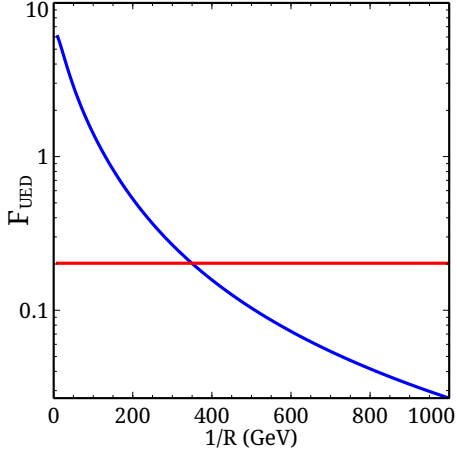}    
    \caption[]{Variation of $F_{\rm UED}$ with $R^{-1}$ in UED model. The horizontal line represents the 95 \% C.L. upper limit on the 
    value of $F_{\rm UED}$ calculated  from the difference between the experimental value of $R_b$ and its theoretical (SM) estimate.}
    \label{feud}
\end{figure} 

 In Fig.\ref{feud}, we plot $F_{\rm UED}$  with $R^{-1}$, the only free parameter in the model after summing contributions ($F_{\rm UED}^{(n)}$) coming from 5 KK-levels  starting from $n =1$. This has been done in the view of recently discovered Higgs mass and its implication on the cut-off scale of UED \cite{explain_vac_st}. Masses of the KK-excitations increase with $R^{-1}$. This in turn set in a decrement of the magnitude of 
$F_{\rm UED}$ due to the higher values of the masses of the propagators in the loop. One can easily check from Eq.\ref{fuedeq}, that
in the limit $r_{1n}^{\prime},  r_{1n} \rightarrow \infty$ , $F_{\rm UED}$ is also vanishing,  telling us the decoupling nature of the theory.  
The horizontal line in Fig.\ref{feud} represents the 95\% C.L. upper limit on the 
 value of $F_{\rm UED}$ calculated  from difference between the experimental value of $R_b$ and its theoretical (SM) estimate 
 ($F_{\rm UED}$: $-0.3165 \pm 0.2647$). So the intersection of the horizontal line 
   with the line showing the variation of $F_{\rm UED}$ would lead us to the present lower bound on $R^{-1}$ from $R_b$. It clearly points that at 95\% C.L. 
   $R^{-1}$ must be greater than 350 GeV, which is a  nominal improvement over the earlier limit which was 300 GeV {\cite{oliver_zbb}. If we ignore the correlation between the Higgs mass and the cut-off scale of UED, then one could sum upto 20-40 levels. This would slightly push up the magnitude of $F_{\rm UED}$\footnote{ For $R^{-1} = 1\rm \;TeV$ values of $F_{\rm UED}$ after summing upto 5 levels and 20 levels are 0.0267 and 0.0292 respectively.}  which in turn results into a higher value of the lower limit of $R^{-1}$ (370 GeV). However, this limit is still not competitive to the bound derived from experimental data on SM Higgs production and its subsequent decay to $WW$ \cite{ayon_ued}\footnote {This is due to the fact that experimental data from LHC on Higgs boson production and subsequent decay to $WW$ is more consistent to the SM than $R_{b}$ in which there is 1.2$\sigma$ new physics window.}.  At this point it would not be irrelevant to discuss very briefly the bounds on $R^{-1}$ in mUED scenario from other experimental data, to put our result into proper context. Constraints from $(g -2)_\mu$ \cite{muon_ued}, FCNC processes \cite{FCNC_ued}, $\rho$-parameter \cite{rho_ued} and other electroweak processes \cite{EW_ued} would result into  $R^{-1} \ge 300$ GeV.  While projected bounds from the LHC is in the ballpark of a TeV \cite{lhc_ued}. However, presently the strongest bound on $R^{-1}$ comes from the consideration of Higgs boson production and decay \cite{ayon_ued} or from the consideration of relic density \cite{belanger}. In the last two cases, the derived limits are comparable and yield $R^{-1} \ge 1.3$ TeV.

\begin{figure}[thb]

 \centering
 \subfigure[]{
   \includegraphics[scale=0.4]{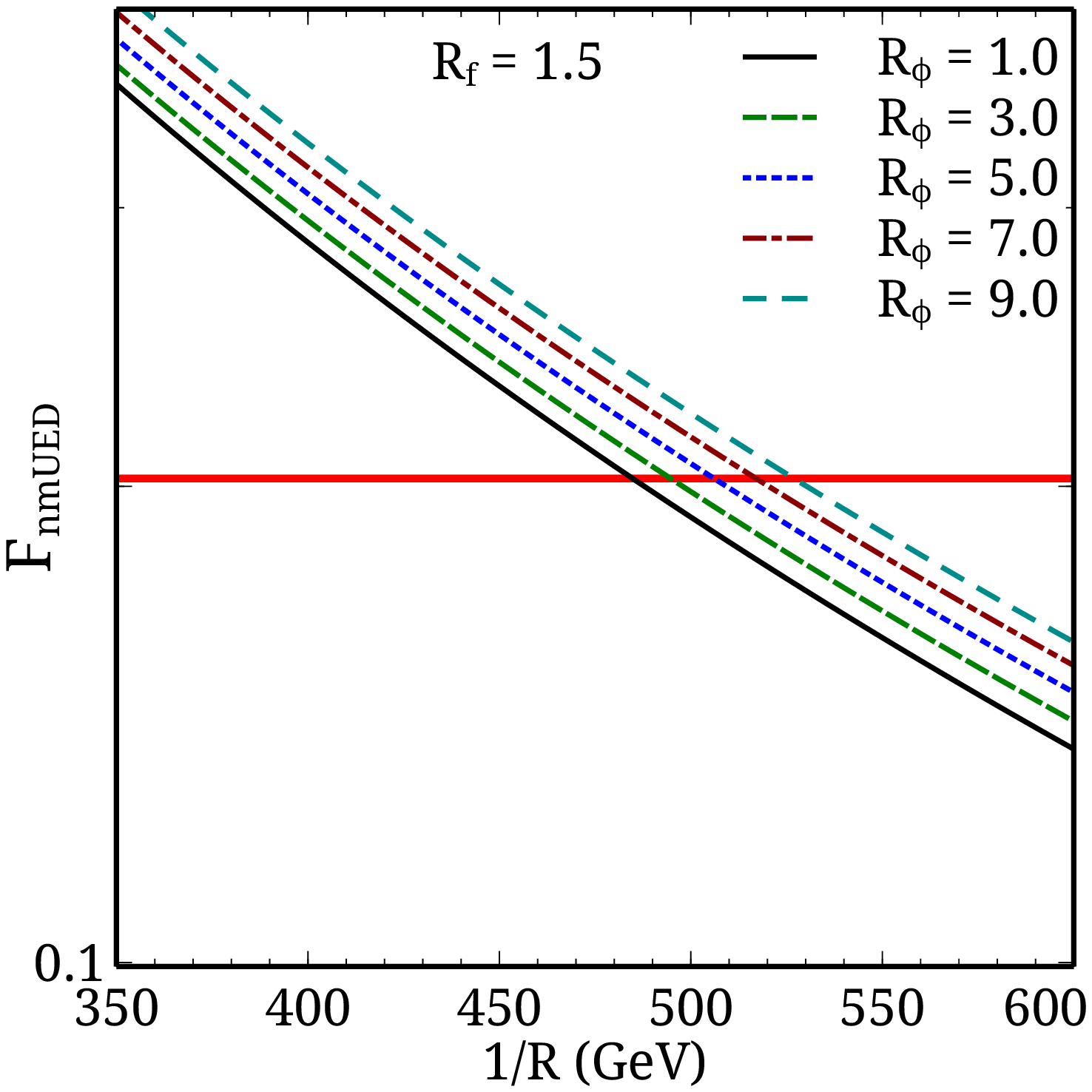}    }
     \subfigure[]{
   \includegraphics[scale=0.4]{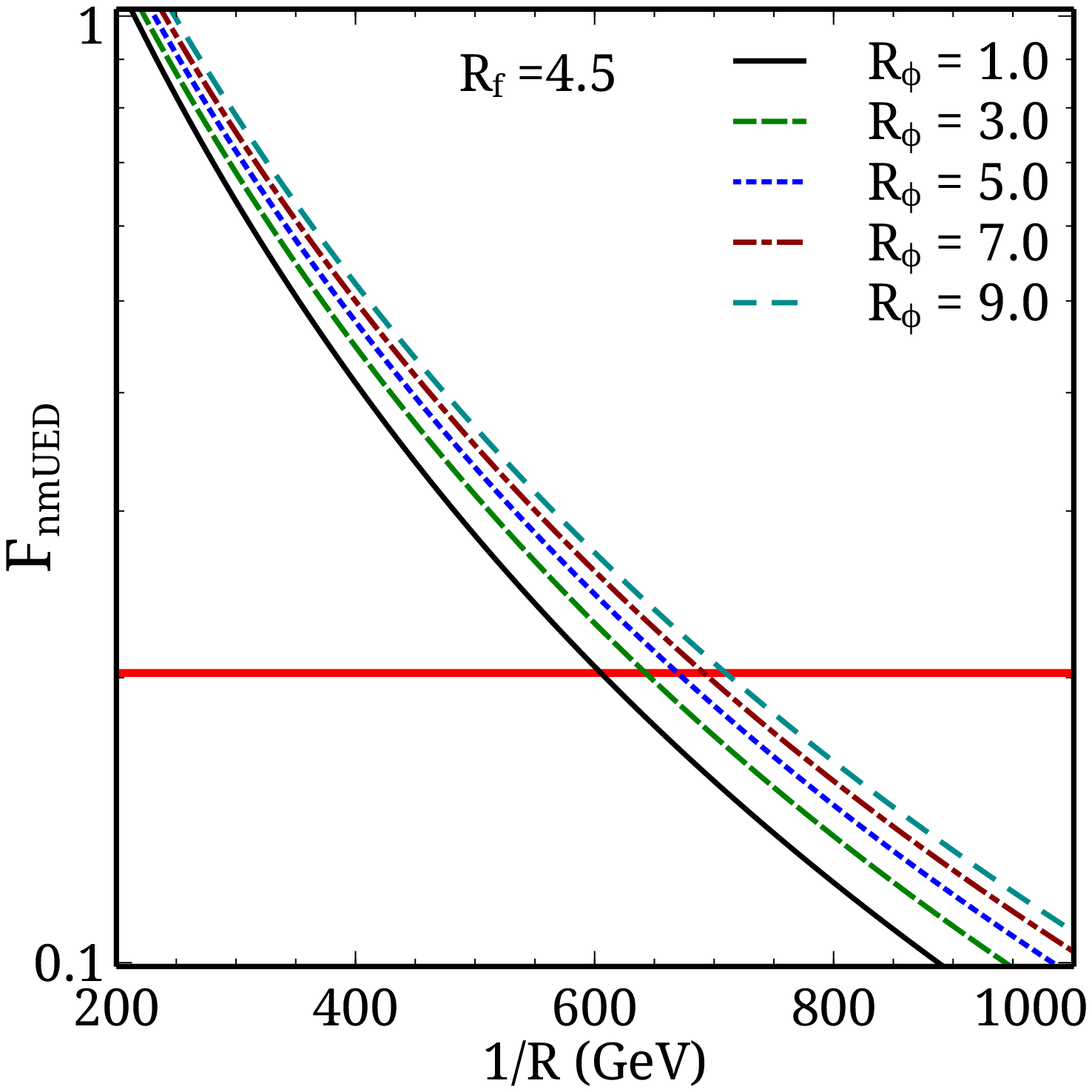}
   } \\
  \subfigure[]{
   \includegraphics[scale=0.4]{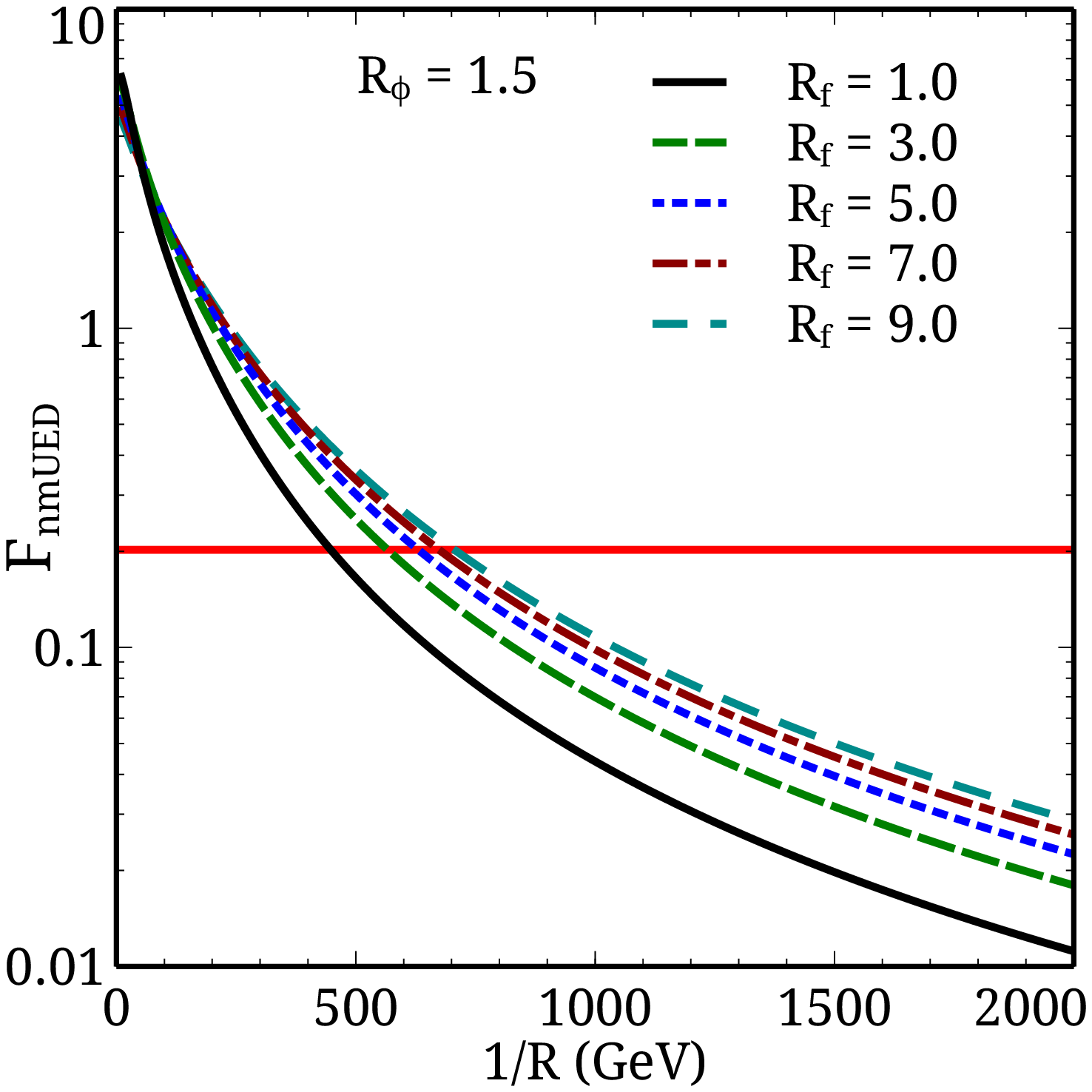}
   } 
  \subfigure[]{
   \includegraphics[scale=0.4]{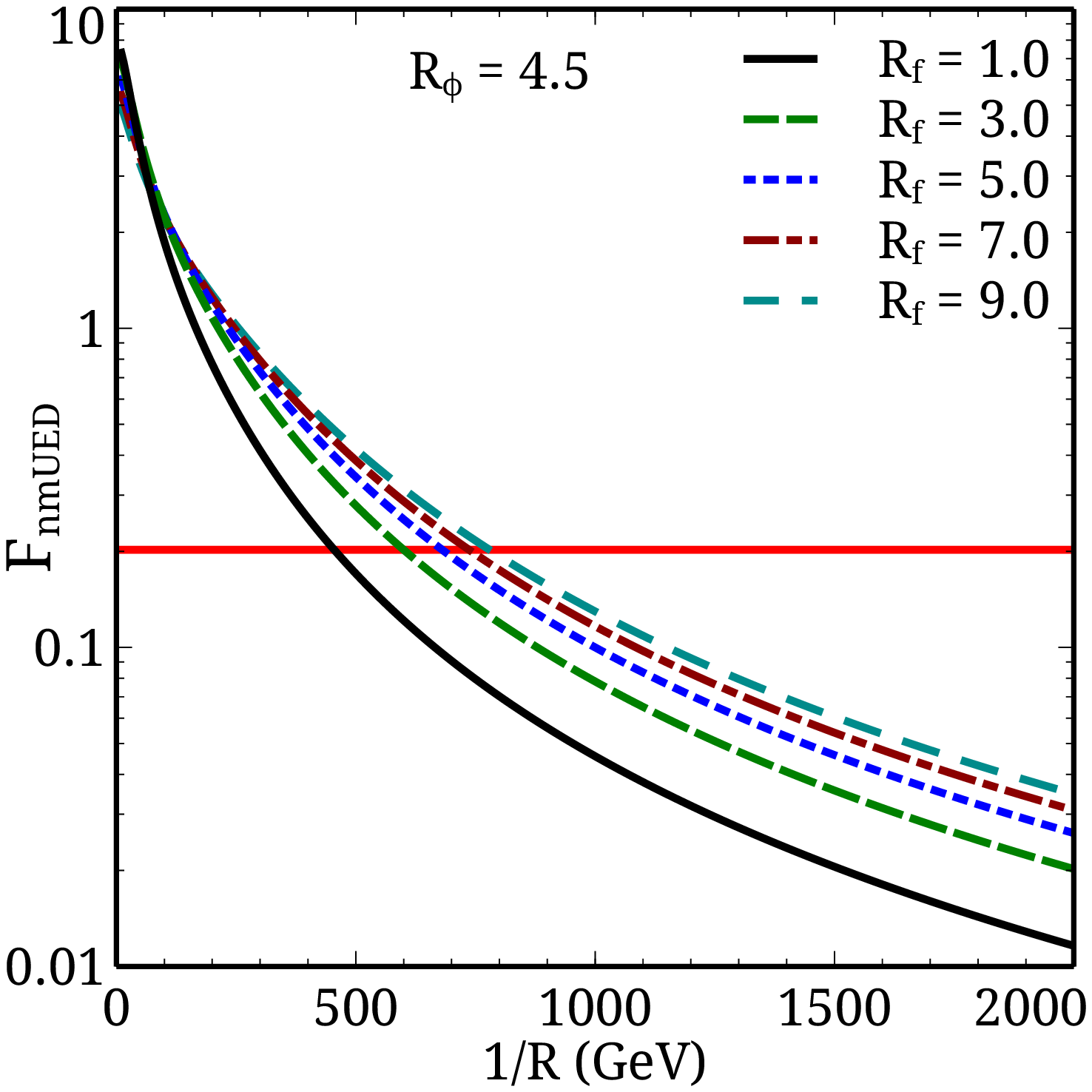}
    }

    \caption[]{Variation of $F_{\rm nmUED}$ with $R^{-1}$ for different values of BLKT parameters. The horizontal line represents the 95 \% C.L. upper limit of the 
    value of $F_{\rm nmUED}$ calculated  from difference between the experimental value of $R_b$ and its theoretical (SM) estimate.}
    \label{fig-onebyr}
\end{figure}

\subsection{Possible bounds on  nmUED from $R_b$} 

Now we are ready to discuss the main results of our analysis. Contribution to $F_{\rm nmUED}$, coming from each KK-level are already listed in the previous section. One has to sum over  KK-levels to get the total contribution. We have taken into consideration the first 5 levels into the summation. And we have explicitly checked that taking 20 levels into the summation, would not  change 
the results\footnote{ For $R^{-1} = 1 \rm \;TeV$ and $r_\phi = 1.5$,  $r_f = 1$, values of $F_{\rm nmUED}$ after summing upto 5 levels and 20 levels are 0.0439 and 0.0472 respectively.}. But before presenting the result, we would like to comment about the values of the BLT parameters used in our analysis. The fermion and the gauge BLKT can be positive or negative. A careful look at eq.\ref{normalisation}
tells us that once the scaled BLT parameters $R_{\phi, f } (\equiv  r_{f, \phi}R^{-1}) < -\pi$, the zero modes become ghost-like {\rm i.e.} its norm becomes imaginary (fermion or gauge). Furthermore, one can also see from eq.\ref{gauge_coup} that  for $R_{\phi}  < -\pi $,  gauge coupling becomes imaginary. So, negative values 
of $R_{f, \phi}$ below $- \pi$ would be physically unacceptable. Apart from this, all other values of $R_{f, \phi}$ are theoretically acceptable. However, one has to be careful in choosing the values of $R_{f,\phi}$, so that the overlap integrals and couplings should not be too large to break the perturbative unitarity. A full analysis imposing unitarity constraints must be carried out to get an idea about this which is beyond the scope of this present article. However, we have restricted ourselves to the values of $R_{f, \phi}$ such that the resulting effective couplings never become 
larger than unity.

\begin{figure}[h]
 \centering
   \includegraphics[scale=0.63]{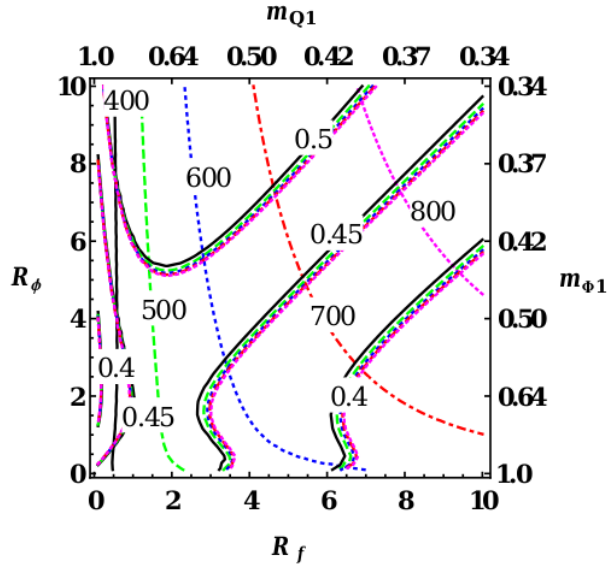}    
    \caption[]{Contours of constant $F_{\rm nmUED}$ corresponding to 95\% C.L. upper limit in $R_{\phi} - R_f$ plane. Different lines (marked with 400, 500, 600, 700 and 800) represent different values of $R^{-1}$. Region right to a particular contour is being ruled out at 95\% C.L. from the consideration of $R_b$ for a given value of $R^{-1}$(in GeV)  on each contour.  We also present contours of the $W^{\pm 1} t^{1} b^{0}$ coupling corresponding to  three different values (0.4, 0.45 and 0.5) on the same plot for a same set of values of $R^{-1}$. Numbers along the top axis and right hand axis correspond to dimensionless quantities $M_{Q 1} R$ and $M_{\Phi {1}} R$   respectively.} \label{cont}
\end{figure}

In Fig.\ref{fig-onebyr}, we have presented the variation of $F_{\rm nmUED}$  as a function of $R^{-1}$ for some representative  values of the {\em scaled} BLKT parameters $R_\phi$ and $R_f$. One common feature that comes out from all of the plots, namely the monotonic decrement of $F_{\rm nmUED}$ with increasing $R^{-1}$, showing the decoupling nature of the new physics under our consideration, which has been pointed out earlier in the case of UED . Panels (a and b) show the dependence of $F_{\rm nmUED}$ on
$R_f$ keeping the value of $R_\phi$  fixed to 1.5 and 4.5 respectively. While in the lower panels of Fig.\ref{fig-onebyr}, we have 
presented how $F_{\rm nmUED}$ changes with varying $R_\phi$ with two fixed values of $R_f$ namely 1.5 (c) and 4.5 (d) respectively.

From the figures it is evident that $R_\phi$ and $R_f$ have more or less same effects on $F_{\rm nmUED}$ and $\delta g_L ^{\rm NP}$ . While the effect of 
$R_\phi$ is  somewhat modest, $F_{\rm nmUED}$ is being more sensitive to any change of $R_f$.  
So by increasing the BLT parameters one could enhance the radiative effects on the effective $Zb\bar{b}$ coupling. 
 Consequently, in nmUED, one could significantly shift  the lower bound on $R^{-1}$, from its UED value. As for example,
for $R_\phi = 1.5$ and $R_f = 9$, the 95 \% C.L. lower bound on $R^{-1}$ is around 700 GeV. This limit comes down to 448 GeV 
for $R_f = 1$.

The role of $R^{-1}$ in the framework of nmUED is similar as in the case of UED and has been explained above. 
We would also like to understand the role of $R_f$ and $R_\phi$.  However we will do so a little later.  

Finally in Fig.\ref{cont}, we present the allowed parameter space in $R_{\phi} - R_f$ plane for several values of $R^{-1}$. We plot the contours of constant $F_{\rm nmUED}$ which corresponds to the 95 \% C.L. upper limit. The region right to a particular line is ruled out from the consideration of $R_{b}$ according to our analysis. Near vertical nature of the contours at lower values of $R_f$ points out to the modest dependence of $F_{\rm nmUED}$ on $R_\phi$ already exhibited in Fig.$4(a)$ and $(b)$. It has been revealed from Fig.\ref{fig-onebyr} that with higher values of BLKT parameters $R_f$ and $R_\phi$, $F_{\rm nmUED}$ is being increased in magnitude. So as we go towards the right with increasing $R_f$ and fixed $R_{\phi}$ for a particular value of $R^{-1}$, $F_{\rm nmUED}$ would increase. Furthermore the higher value of $R^{-1}$ decreases $F_{\rm nmUED}$ showing the decoupling nature of new physics. Thus the increment of $F_{\rm nmUED}$ (with $R_f$) has been nullified by higher values of $R^{-1}$ corresponding to different lines. Thus to compensate one must tune $R^{-1}$ to a comparatively higher values.  Furthermore, we have marked axes of Fig.\ref{cont} with scaled masses $m_{Q 1}$ ($\equiv M_{Q1} R)$
and  $m_{\Phi 1}$ ($\equiv M_{\Phi 1} R$). This would facilitate one to read off the bounds on masses of the $n =1$ KK-excitations 
directly from this plot. As for example, the line corresponding to $R^{-1} = 700$ GeV intersects, the $M_{Q1}$ axis at 
around 0.5, which implies, that for this particular value of $R^{-1}$, masses of $n=1$ KK-excitations of top quarks below
350 Gev are not allowed by the data. While the corresponding lower bound for $W^1$ mass for $R^{-1} = 700$ GeV
is close to 560 GeV which can be read off from the intersection of the same line with the $m_{\Phi 1}$ axis.  

In Fig.\ref{cont}, contours for constant (for three different) values of $W^{\pm 1} f^{1} f^0$ couplings have also been presented. One can 
get several important messages from these contours. Primarily the above coupling has a minimal dependence on $R^{-1}$.   Secondly, 
BLT parameters  $R_f$ and $R_\phi$ have opposite effects on the above interaction. While this coupling increases with $R_\phi$, increasing values of $R_f$ would try to decrease the  strength of this interaction. Similar 
conclusion can be drawn to  $H^{\pm 1} f^{1} f^0$ and $G^{\pm 1} f^{1} f^0$ interactions.  BLT parameters have another bearing on $F_{\rm nmUED}$ through the masses of KK-excitations. Heavier KK-masses would tend to decrease the magnitude of $F_{\rm nmUED}$.
It has been pointed out earlier that KK-masses are decreasing function of respective BLT parameters.  Thus BLT parameters have dual role
to play in the dynamics of $F_{\rm nmUED}$.  Let us state them one by one. An increasing $R_\phi$ would increase $F$ by increasing the relevant couplings and at the same time by decreasing the relevant KK-masses. On the other hand an increasing $R_f$  would decrease the masses but it also decreases the couplings. These two effects play in opposite direction in determining the value of $F$. However, rate at    which $F$ increases with decreasing KK-mass,  overcome the decrement of $F$ due to decreasing coupling (with increasing $R_f$).

Before we conclude, we would like to comment on the terms which we have neglected by only considering interactions of
 SM particles with two KK-excitations of same KK-number. As a consequence of these our calculation and results presented above
 do not take into account a number of Feynman graphs in which propagators in the loop corresponds to KK-excitations of different 
 KK-numbers.  To advocate our assumption,  we present the values of $F_{\rm nmUED}$ for several vales of $R^{-1}$
 and for fixed values of $R_\phi$ and $R_{f}$ in Table.\ref{table}.  While presenting these numbers we have summed upto $5$ KK-levels as before. 
 In the second column of Table.\ref{table} we present the values of $F_{\rm nmUED}$ when only KK-number conserving interactions are taken into account. While in the third column, values of $F_{\rm nmUED}$ correspond to the case when all possible Feynman graphs involving KK-number violating interactions have been taken into account. It is evident from the numerical values of $F_{\rm nmUED}$ that our assumption was 
 realistic and the corrections coming from Feynman graphs involving the KK-number non-conserving interactions are minuscule.

\begin{table}[h]
\begin{center}
\begin{tabular}{|c|c|c|} 
\hline 
$R^{-1}$ (GeV) & \begin{tabular}[c]{@{}c@{}}$F_{\rm nmUED}$\\ ($n=m$ terms only)\end{tabular} & \begin{tabular}[c]{@{}c@{}}$F_{\rm nmUED}$\\ (all interactions )\end{tabular} \\ \hline
250         & 0.5442                                                     & 0.5481                                                               \\ \hline
350         & 0.3127                                                     & 0.3148                                                               \\ \hline
450         & 0.2003                                                      & 0.2016                                                                \\ \hline
550         & 0.1384                                                      & 0.1393                                                               \\ \hline
650         & 0.1009                                                      & 0.1016                                                                \\ \hline
750         & 0.0767                                                      & 0.0773                                                                \\ \hline
850         & 0.0602                                                      & 0.0606                                                                \\ 
\hline
\end{tabular}
\caption{Values of $F_{\rm nmUED}$ when only the KK-number conserving interactions are taken (second column) 
and when all possible interactions are taken into account (third column) to calculate the effective $Zb\bar{b}$ vertex at one loop. Numbers are presented for several values of $R^{-1}$ (first column) and for $R_{f}$=1 and $R_{\phi}$=1.5.}
\label{table}
\end{center}
\end{table}


 \section{Conclusion}
 In summary, we have estimated one loop contribution to the $Zb\bar{b}$ vertex in the framework of an Universal Extra Dimensional Model where kinetic and Yukawa terms are added to the fixed points of the extra space like dimension. These boundary-localized terms, with their coefficients as free parameters, parametrize the quantum corrections to the masses of the KK-excitations and their mutual interactions. We have calculated the interactions necessary for our calculation. Some of these interactions are very similar to those in UED. However, some of the interactions are modified in comparison to their UED counterparts by some overlap integrals (in extra dimension) involving the extra dimensional profiles of the fields present in an interaction vertex.
 
 The effect of BLKTs on the masses of KK-modes and their interactions can be summarized  as the following. For a given $R^{-1}$, increasing BLKT parameter would drive the respective masses to lower values. Strength of an interaction does not have such a simple dependence on the BLKT parameters. We have derived all the necessary interactions involving the KK-excitations of top quarks, W bosons, charged Higgs and Goldstone bosons in the framework of nmUED with the assumption of equal gauge and Higgs BLKT along with equal fermion and Yukawa BLKTs. Gauge and Higgs BLKTs have been chosen to be equal to avoid the unnecessary complication created by the presence of $r_\phi$ in  equation of motion of gauge fields. While unequal fermion BLKTs and Yukawa BLT would lead to the KK-mode mixing in the definition of physical states of KK-excitations of top quarks. So for the sake of a relatively simpler calculation we stick to the choice of  equal fermion BLKTs and Yukawa BLTs.

  In general, coupling of a $b$ quark to the $Z$ boson involves both the left- and right-chiral projectors. However, quantum corrections which go into the coefficient of the left-chiral projector are proportional to $m_t^{2}$ while the $m_b^{2}$ proportional terms go into the coefficient of the right-chiral projector. We have done the calculation in the limit where $m_b \rightarrow 0$.  There are two main classes of Feynman diagrams contributing to $\delta g_L ^{\rm NP}$, (the contribution to $Zb\bar{b}$ vertex in nmUED framework) in 't-Hooft Feynman gauge. First set of diagrams listed in Fig.\ref{fig1}, captures the dominant contribution (because of Yukawa coupling which is proportional to $m_t$) coming from the participation of KK-excitations of top quarks and charged Higgs boson/Goldstones in the loops. 
 The remaining set consists of contribution mainly coming from the KK-excitations of $W$ bosons and top quarks inside the loops. These diagrams are listed in Fig.\ref{fig2}.
 
 The explicit expressions for the contributions coming from each of the diagrams are listed in the section 3.  Sum of the contributions to $\delta g_L ^{\rm NP}$ from the diagrams in Fig.\ref{fig1} is finite and independent of $\sin^2 \theta_W$. While the second set of diagrams needs to be regularized,  after summing up, it is still ultraviolet divergent and also contains a term which grows with $R^{-1}$.\footnote{This term arises from the diagram $2(f)$, due to a direct proportionality on $R^{-1}$ of the vertex $W-G-Z$.} We have used a regularisation scheme following Ref.\cite{pichsanta, buras}, upon which the total contribution becomes finite and also becomes independent of $\sin^2 \theta_W$.
 
A recent theoretical estimation of the $Zb\bar{b}$ vertex in the framework of SM at two loop level has squeezed the window for new physics that might be operating at TeV scale. The experimentally measured value of $R_b$ differs from the SM prediction at 1.2 $\sigma$ level. We have used the experimental data and the recent results from the SM on $R_b$, to constrain the parameter space of non-minimal Universal Extra Dimensional Model. We have relooked into the UED by setting the BLKT parameters to zero in our calculation. The resulting expressions can be used to put bound on $R^{-1}$ in UED model using the same experimental data and the SM estimations of $R_b$. It has been found that $R^{-1}$ in UED model should be greater than 350 GeV at 95 \% C.L.

Next we focus into our main result. Comparing the numerical estimation of $F_{\rm nmUED}$ with the difference between experimental data and SM estimation we have constrained the parameters in nmUED. First we look into the limits on $R^{-1}$. Both the BLKT parameters have positive effects on $F_{\rm nmUED}$. This function is very sensitive to any change in $R_f$ while the effect of
$R_\phi$ is very mild. The bottom line is that both the BLKT parameters can push the allowed value of $R^{-1}$ to higher values. Depending on magnitude of BLKT parameters $R_{\phi}$ and $R_f$ (which we have chosen to be positive), lower limit on $R^{-1}$ could be close to 800 GeV. Finally, we plot contours of constant $F_{\rm nmUED}$ having the 95 \% C.L. upper limit value for different value of $R^{-1}$ in $R_{\phi}-R_{f}$ plane. As for a fixed value of $R^{-1}$ i.e. for a fixed curve the value of the function $F_{\rm nmUED}$ increases with increase of $R_{f}$ the left side of that curve represents the allowed region of this function for respective $R^{-1}$.

{\bf Acknowledgements:}  AD acknowledges partial financial support from BRNS-DAE, Govt. of India. TJ acknowledges financial support from 
CSIR in terms of a JRF. Both the authors are grateful to Gautam Bhattacharyya, Amitava Raychaudhuri and Avirup Shaw for many useful discussions.

%

%
%
%
%
%
%

\vspace{1cm}
\begin{center}
{\bf \large APPENDIX A}
\end{center}
In the following Feynman rules, all momenta and fields are assumed to flow into the vertices.

\begin{figure}[h]
  \includegraphics[scale=0.7]{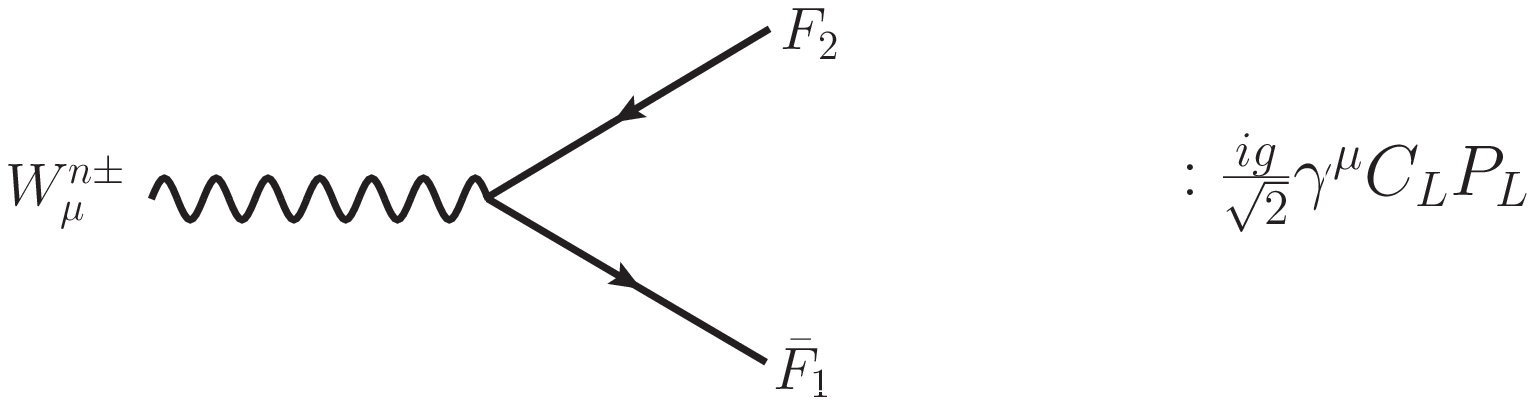}
\end{figure}

\begin{eqnarray*}
W^{n +}\bar{Q}_{t}^{\prime n}b_{L}^{0}  \colon   C_{L} = -I_{1}\sqrt \beta\cos\alpha_{n},
                                       ~~~~~~~~~~~~~~~~~~~~~~~~~ W^{n -}\bar{b}_{L}^{0}Q_{t}^{\prime n}  \colon  
                                      C_{L} = -I_{1}\sqrt \beta\cos\alpha_{n},\\
W^{n +}\bar{U}^{\prime n}b_{L}^{0}  \colon   C_{L} = I_{1}\sqrt \beta\sin\alpha_{n},
                                       ~~~~~~~~~~~~~~~~~~~~~~~~~~~~~ W^{n -}\bar{b}_{L}^{0}U^{\prime n}  \colon  
                                      C_{L} = I_{1}\sqrt \beta\sin\alpha_{n}.
\end{eqnarray*}

\begin{figure}[h]
  \includegraphics[scale=0.7]{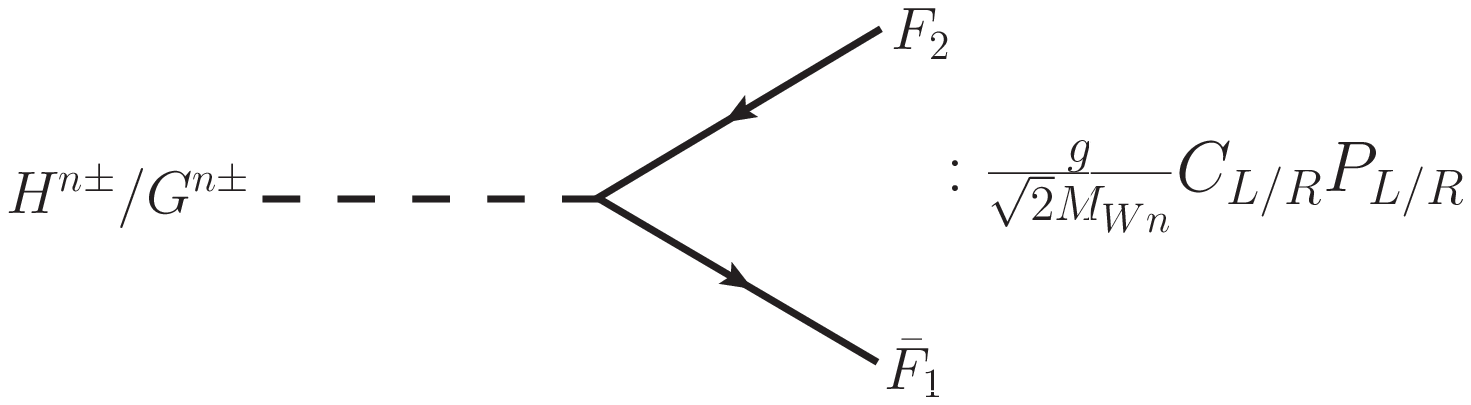}
\end{figure}
\begin{flushleft}
\begin{eqnarray*}
H^{n +}\bar{Q}_{t}^{\prime n}b_{L}^{0}  \colon   C_{L} &=& -i\sqrt \beta\left(I_{1}\frac{m_{t}M_{\Phi n}}{M_{W}}\sin\alpha_{n} - I_{2}M_{W}   \cos\alpha_{n}\right),\\
H^{n -}\bar{b}_{L}^{0}Q_{t}^{\prime n}  \colon   C_{R} &=& -i\sqrt \beta\left(I_{1}\frac{m_{t}M_{\Phi n}}{M_{W}}\sin\alpha_{n} - I_{2}M_{W}
\cos\alpha_{n}\right),\\
G^{n +}\bar{Q}_{t}^{\prime n}b_{L}^{0}  \colon   C_{L} &=& \sqrt \beta\left(I_{1}m_{t}\sin\alpha_{n} + I_{2}M_{\Phi n}\cos\alpha_{n}\right),\\
G^{n -}\bar{b}_{L}^{0}Q_{t}^{\prime n}  \colon   C_{R} &=& -\sqrt \beta\left(I_{1}m_{t}\sin\alpha_{n} + I_{2}M_{\Phi n}\cos\alpha_{n}\right),\\
H^{n +}\bar{U}^{\prime n}b_{L}^{0}  \colon   C_{L} &=& i\sqrt \beta\left(I_{1}\frac{m_{t}M_{\Phi n}}{M_{W}}\cos\alpha_{n} + I_{2}M_{W}
\sin\alpha_{n}\right), \\
H^{n -}\bar{b}_{L}^{0}U^{\prime n}  \colon   C_{R} &=& i\sqrt \beta\left(I_{1}\frac{m_{t}M_{\Phi n}}{M_{W}}\cos\alpha_{n} + 
                                                                 I_{2}M_{W}\sin\alpha_{n}\right),\\
G^{n +}\bar{U}^{\prime n}b_{L}^{0}  \colon   C_{L} &=& -\sqrt \beta\left(I_{1}m_{t}\cos\alpha_{n} - I_{2}M_{\Phi n}\sin\alpha_{n}\right),\\
G^{n -}\bar{b}_{L}^{0}U^{\prime n}  \colon   C_{R} &=&\sqrt \beta\left(I_{1}m_{t}\cos\alpha_{n} - I_{2}M_{\Phi n}\sin\alpha_{n}\right).
\end{eqnarray*}
\end{flushleft}

\begin{figure}[h]
  \includegraphics[scale=0.7]{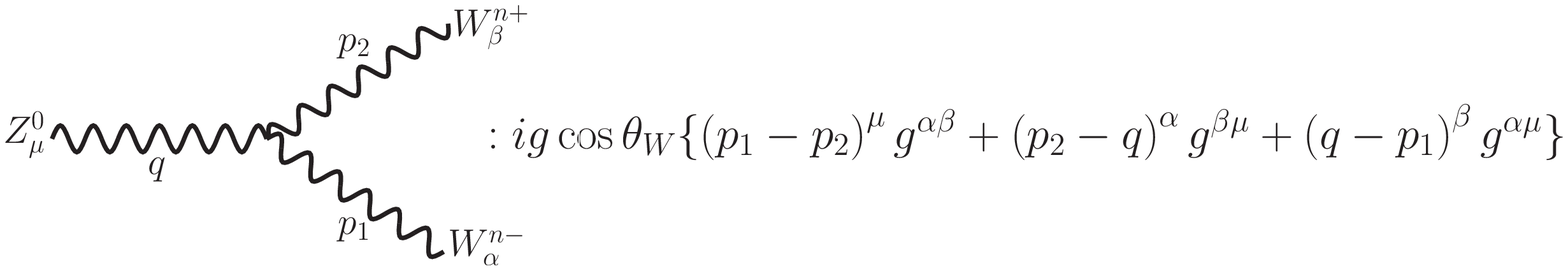}
\end{figure}

\begin{figure}[h]
  \includegraphics[scale=0.7]{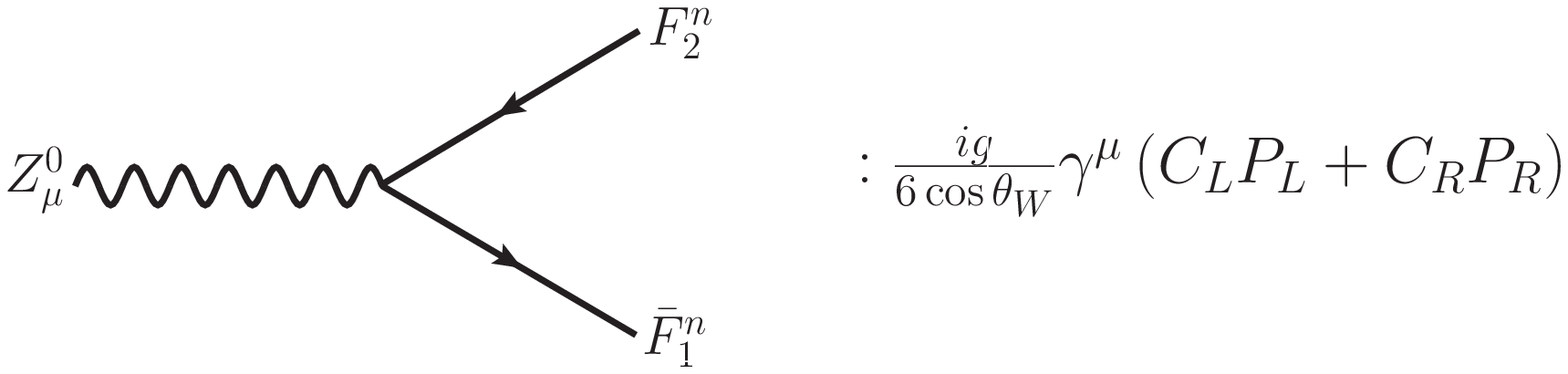}
\end{figure}
\begin{equation*}
Z^{0}\bar{Q}_{t}^{\prime n}Q_{t}^{\prime n}  \colon  \left\{ \begin{array}{rl}
                                      C_{L} = -4\sin^{2}\theta_{W}+3{\cos}^{2}\alpha_{n} \\
                                      C_{R} = -4\sin^{2}\theta_{W}+3{\cos}^{2}\alpha_{n}
                                      \end{array} \right., ~~~ Z^{0}\bar{U}^{\prime n}U^{\prime n}  \colon  \left\{ \begin{array}{rl}
                                      C_{L} = -4\sin^{2}\theta_{W}+3{\sin}^{2}\alpha_{n} \\
                                      C_{R} = -4\sin^{2}\theta_{W}+3{\sin}^{2}\alpha_{n}
                                     \end{array} \right.,\\
\end{equation*}
\begin{equation*}
Z^{0}\bar{Q}_{t}^{\prime n}U^{\prime n}  \colon  \left\{ \begin{array}{rl}
                                      C_{L} = -3\sin\alpha_{n}\cos\alpha_{n} \\
                                      C_{R} = -3\sin\alpha_{n}\cos\alpha_{n}
                                      \end{array} \right., ~~~ Z^{0}\bar{U}^{\prime n}Q_{t}^{\prime n}  \colon  \left\{ \begin{array}{rl}
                                      C_{L} = -3\sin\alpha_{n}\cos\alpha_{n} \\
                                      C_{R} = -3\sin\alpha_{n}\cos\alpha_{n}
                                      \end{array} \right..
\end{equation*}

\begin{figure}[h]
  \includegraphics[scale=0.7]{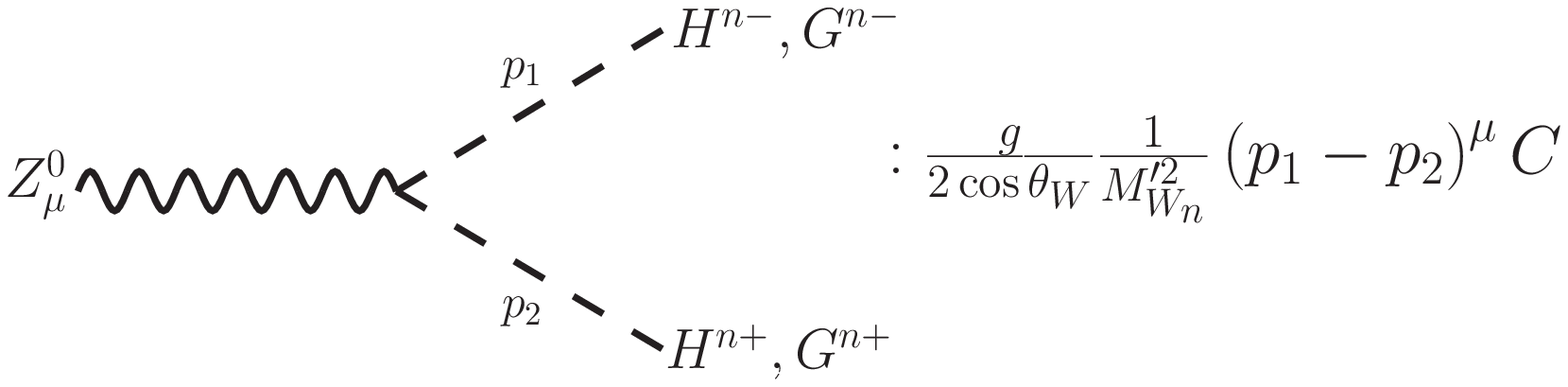}
\end{figure}
\begin{flushleft}
\begin{eqnarray*}
Z^{0}H^{n +}H^{n -}  \colon  C &=& i\{(-1+2\sin^{2}\theta_{W})M_{\Phi n}^{2} - 2\cos^{2}\theta_{W}M_{W}^{2}\},\\
Z^{0}G^{n +}G^{n -}  \colon  C &=& i\{(-1+2\sin^{2}\theta_{W})M_{W}^{2} - 2\cos^{2}\theta_{W}M_{\Phi n}^{2}\},\\
Z^{0}H^{n -}G^{n +}  \colon  C &=& -M_{\Phi n}M_{W},\\
Z^{0}G^{n -}H^{n +}  \colon  C &=& M_{\Phi n}M_{W}.
\end{eqnarray*}
\end{flushleft}
\newpage
\begin{figure}[h]
  \includegraphics[scale=0.7]{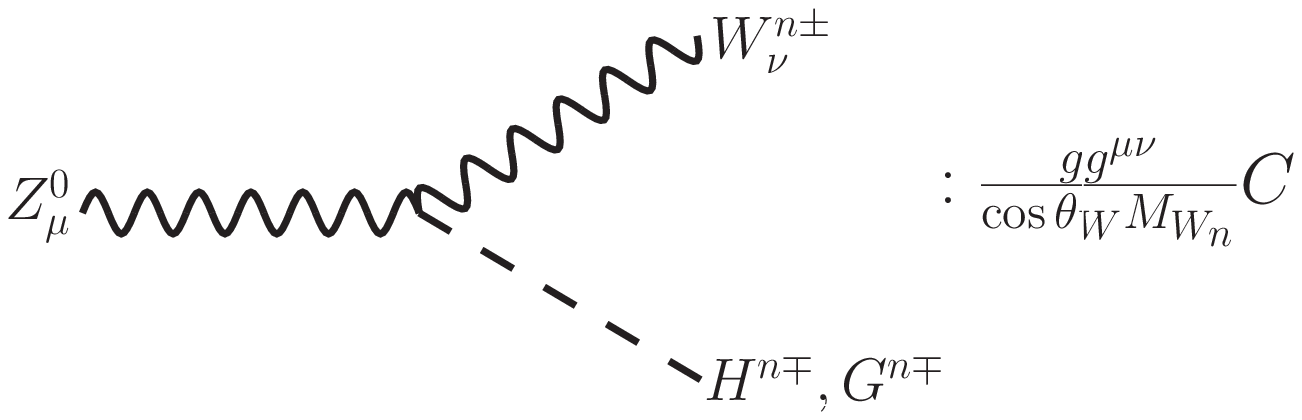}
\end{figure}
\begin{flushleft}
\begin{eqnarray*}
Z^{0}W^{n +}G^{n -}  \colon  C &=& \left(-M_{W}^{2}\sin^{2}\theta_{W} + M_{\Phi n}^{2}\cos^{2}\theta_{W}\right),\\
Z^{0}W^{n -}G^{n +}  \colon  C &=& \left(M_{W}^{2}\sin^{2}\theta_{W} - M_{\Phi n}^{2}\cos^{2}\theta_{W}\right),\\
Z^{0}W^{n +}H^{n -}  \colon  C &=& -iM_{\Phi n}M_{W},\\
Z^{0}W^{n -}H^{n +}  \colon  C &=& -iM_{\Phi n}M_{W}.
\end{eqnarray*}
\end{flushleft}
$I_{1}$ and $I_{2}$ have the following form:
\begin{eqnarray*}
I_{1} &=& \frac{2}{\pi R}\left[ \frac{1}{\sqrt{1 + \frac{r_f^2 M_{Qn}^{2}}{4} + \frac{r_f}{\pi R}}}\right]\left[ \frac{1}{\sqrt{1 + \frac{r_\phi^2 M_{\Phi n}^{2}}{4} + \frac{r_\phi}{\pi R}}}\right]\frac{M_{\Phi n}^{2}\left(-r_{f} + r_{\phi}\right)}{\left(M_{Q n}^{2} - M_{\Phi n}^{2}\right)},\\
I_{2} &=& \frac{2}{\pi R}\left[ \frac{1}{\sqrt{1 + \frac{r_f^2 M_{Qn}^{2}}{4} + \frac{r_f}{\pi R}}}\right]\left[ \frac{1}{\sqrt{1 + \frac{r_\phi^2 M_{\Phi n}^{2}}{4} + \frac{r_\phi}{\pi R}}}\right]\frac{M_{\Phi n} M_{Q n}\left(-r_{f} + r_{\phi}\right)}{\left(M_{Q n}^{2} - M_{\Phi n}^{2}\right)}.
\end{eqnarray*}

\end{document}